\documentclass[aps,twocolumn, amssymb,amsmath,floatfix,superscriptaddress,longbibliography]{revtex4-1}
\pdfoutput=1

\usepackage[colorlinks=true,citecolor=blue,linkcolor=magenta]{hyperref}
\usepackage{amsmath}
\usepackage{graphicx}
\usepackage{soul}
\usepackage{changepage}
\usepackage{fancyhdr}
\usepackage{amsthm,amssymb}
\usepackage{floatflt}
\usepackage{float}
\usepackage{array}
\usepackage[usenames,dvipsnames]{color}
\usepackage{appendix}
\usepackage[export]{adjustbox}

\newcommand{\bea}{\begin{eqnarray}}  
\newcommand{\eea}{\end{eqnarray}}
\newcommand{\ben}{\begin{enumerate}}
\newcommand{\een}{\end{enumerate}}
\newcommand{\be}{\begin{equation}}
\newcommand{\ee}{\end{equation}}

\renewcommand{\theequation}{\arabic{section}.\arabic{equation}}

\newcommand{\norm}[1]{\left\|#1\right\|}

\newcommand{\bra}[1]{\left\langle#1\right|}
\newcommand{\ket}[1]{\left|#1\right\rangle}

\newcommand{\adj}[1]{#1^{\dagger}}

\DeclareMathOperator{\Tr}{Tr}

\newcommand{\wS}{\mathcal{F}}
\newcommand{\vc}[1]{\boldsymbol{#1}}
\newcommand{\avg}[1]{\langle #1 \rangle}

\begin{document}

\title{Integrable and chaotic dynamics of spins coupled to an optical cavity}

\author{Gregory Bentsen}
\thanks{These authors contributed equally to the paper.}
\affiliation{Department of Physics, Stanford University, Stanford, CA 94305, USA}
\affiliation{SLAC National Accelerator Laboratory, Menlo Park, California 94025, USA}

\author{Ionut-Dragos Potirniche}
\thanks{These authors contributed equally to the paper.}
\affiliation{Department of Physics, University of California, Berkeley, Berkeley, CA 94720, USA}

\author{Vir B. Bulchandani}
\affiliation{Department of Physics, University of California, Berkeley, Berkeley, CA 94720, USA}

\author{Thomas Scaffidi}
\affiliation{Department of Physics, University of California, Berkeley, Berkeley, CA 94720, USA}
\affiliation{Department of Physics, University of Toronto, Toronto, Ontario, M5S 1A7, Canada}

\author{Xiangyu Cao}
\affiliation{Department of Physics, University of California, Berkeley, Berkeley, CA 94720, USA}

\author{Xiao-Liang Qi}
\affiliation{Department of Physics, Stanford University, Stanford, CA 94305, USA}

\author{Monika Schleier-Smith}
\affiliation{Department of Physics, Stanford University, Stanford, CA 94305, USA}
\affiliation{SLAC National Accelerator Laboratory, Menlo Park, California 94025, USA}

\author{Ehud Altman}
\affiliation{Department of Physics, University of California, Berkeley, Berkeley, CA 94720, USA}
\affiliation{Materials Science Division, Lawrence Berkeley National Laboratory, Berkeley, CA 94720, USA}

\date{\today}
\begin{abstract}
We show that a class of random all-to-all spin models, realizable in systems of atoms coupled to an optical cavity, gives rise to a rich dynamical phase diagram due to the pairwise separable nature of the couplings. By controlling the experimental parameters, one can tune between integrable and chaotic dynamics on the one hand, and between classical and quantum regimes on the other hand. For two special values of a spin-anisotropy parameter, the model exhibits rational-Gaudin type integrability and it is characterized by an extensive set of spin-bilinear integrals of motion, independent of the spin size. More generically, we find a novel integrable structure with conserved charges that are not purely bilinear. Instead, they develop ``dressing tails'' of higher-body terms, reminiscent of the dressed local integrals of motion found in Many-Body Localized phases. Surprisingly, this new type of integrable dynamics found in finite-size spin-1/2 systems disappears in the large-$S$ limit, giving way to classical chaos. We identify parameter regimes for characterizing these different dynamical behaviors in realistic experiments, in view of the limitations set by cavity dissipation. 
\end{abstract}

\maketitle

\begin{center}
\begin{figure}[t]
\includegraphics[width=\columnwidth]{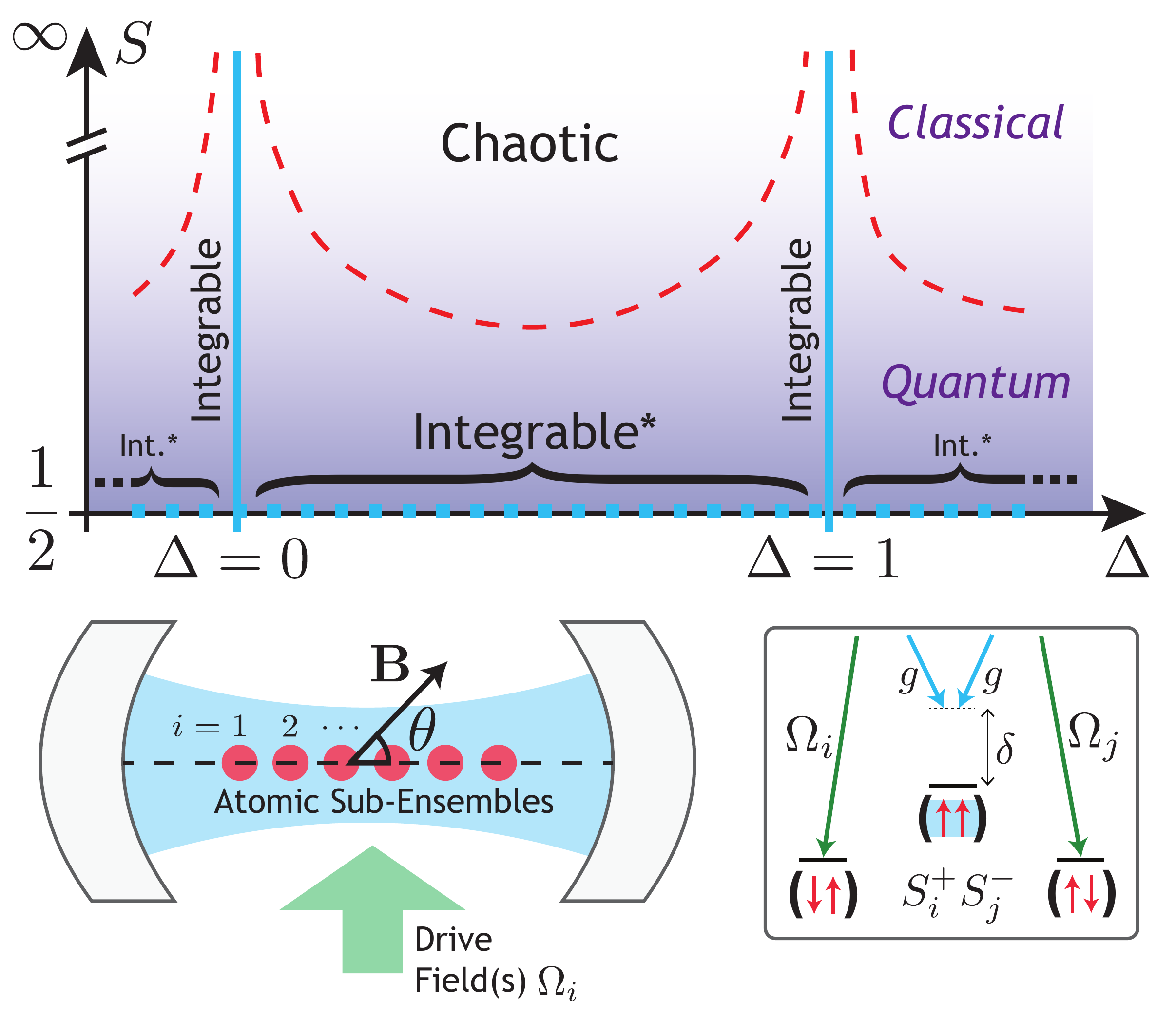}
\caption{(top) The dynamical phase diagram of the model~\eqref{eq:SeparableModelDelta} in the plane of the spin size $S$ and anisotropy $\Delta$. The main features are: the integrable lines at $\Delta=0$ and $\Delta=1$ (solid blue); the novel integrable line at $S=1/2$ (dotted blue and marked by an asterisk); and the onset of chaos at large $S$, indicated schematically by the dashed red curves.\\
(bottom) Schematic of the atomic sub-ensembles (red) trapped inside a single-mode optical cavity (blue). A drive field (green) at a detuning $\delta$ from the cavity resonance generates effective spin-spin interactions between the atoms (bottom right). The tunable angle $\theta$ between the spin quantization axis (along the applied magnetic field $\mathbf{B}$) and the cavity's longitudinal axis leads to an anisotropy $\Delta = 2 \cot^2 \theta$. By changing the local atomic density in a region of constant coupling to the cavity mode, the effective spin size $S$ can also be varied, allowing for the systematic exploration of the full phase diagram.}
\label{fig:PhaseDiagram}
\end{figure}
\end{center}

\section{Introduction}
\label{sec:Introduction}

The dynamics leading to the eventual thermalization of closed quantum systems has become a topic of intense interest over the past few years. Significant progress has been made in describing the scrambling of  information through quantum chaos, which allows effectively irreversible dynamics to emerge from unitary quantum time evolution. Notably, Maldacena \emph{et al.}, inspired by the chaotic properties of black holes, established that quantum mechanics places an upper bound on the Lyapunov exponent that characterizes the growth of chaos~\cite{Maldacena2016bound}. In a related development, Kitaev constructed a class of quantum many-body models whose dynamics saturates this bound on chaos~\cite{Kitaev2015,Stanford} and can be related to black holes through the AdS/CFT correspondence~\cite{Sachdev2015,Jensen2016,Witten2017}. The fact that these models admit controlled solutions, despite being chaotic, has further conferred on them a paradigmatic status within the field of quantum dynamics. 

Finding accessible systems which realize such models is therefore highly desirable, but also, \emph{a priori}, very challenging: a common feature shared by all of these maximally chaotic, holographic models is that they lack spatial locality, since they couple together an extensive number of degrees of freedom. For instance, the Sachdev-Ye (SY) model~\cite{Sachdev1992} was originally proposed as a quantum spin model with random all-to-all couplings:
\begin{equation}
    H = \frac{1}{\sqrt{NM}}\sum_{i,j=1}^{N} U_{ij} \mathbf{S}_i\cdot\mathbf{S}_j,
    \label{eq:SY}
\end{equation}
where $\mathbf{S}_i$ are $\mathrm{SU}(M)$ spin operators. A fermionic variant, the Sachdev-Ye-Kitaev (SYK) model, was subsequently introduced by Kitaev.

While infinite-range spin interactions do not occur in magnetic materials, they can be realized rather naturally in cold atomic ensembles coupled to an optical cavity mode~\cite{Black2003,Majer2007,leroux2010implementation,vanLoo1494,Barontini1317,Hosten2016science,Kollar2017,Leonard2017,norcia2018cavity,Kroeze2018,Landini2018,kohler2018negative,Guo2018,davis2019photon,Braverman2019}. In this setup, the delocalized cavity mode mediates infinite-range interactions between the internal states of atoms through the local coupling at each site, regardless of the distance between atoms~\cite{Andre2002,Sorensen2002,Leroux2010theory,gopalakrishnan2011frustration,HungE4946,Masson2017,Mivehvar2017,Colella2018,Mivehvar2019}. However, there is a crucial difference, already pointed out in Ref.~\onlinecite{Marino2018}, between the interactions in the SY model and the ones mediated by the cavity. The second-order process that couples the atomic degrees of freedom via the cavity mode gives a separable (rank-$1$) matrix $U_{ij}=J_i J_j$, rather than the full-rank matrix assumed in the different variants of the SY model. Although non-separable interactions are, in principle, accessible in multi-mode cavities~\cite{gopalakrishnan2011frustration,strack2011dicke,swingle2016measuring,Kollar2017,Leonard2017,Marino2018,Guo2018}, separable all-to-all couplings are realized in numerous existing experiments~\cite{leroux2010implementation,Hosten2016science,Kroeze2018,Landini2018,norcia2018cavity,davis2019photon,Braverman2019,kohler2018negative} and arise generically for interactions mediated by a single bosonic mode.

Moreover, this ostensible limitation of the cavity-QED scheme turns out to be a boon: the separability of the interaction is responsible for an even richer dynamical phase diagram (see Fig.~\ref{fig:PhaseDiagram}), which includes regions of chaos, Gaudin-type integrability characterized by spin-bilinear conserved quantities, and of a novel form of integrability---labeled Integrable$^*$ in Fig.~\ref{fig:PhaseDiagram}---with \textit{quasi}-bilinear integrals of motion.

The class of models we consider in this paper is described by the following quantum spin Hamiltonian:
\begin{equation}
    H = \frac{1}{S\sqrt{N}}\sum_{i,j=1}^{N} J_i J_j \left(S_i^x S_j^x + S_i^y S_j^y + \Delta S_i^z S_j^z \right),
    \label{eq:SeparableModelDelta}
\end{equation}
where $\mathbf{S}_i$ are $\mathrm{SU}(2)$ spin-$S$ operators encoded in the magnetic sub-levels of individual atoms or atomic sub-ensembles located at sites $i = 1,\dots,N$. The site-dependent coefficients $J_i$ are determined by the local coupling of the atoms at site $i$ to the spatially-varying cavity mode, or by the local Rabi frequency $\Omega_i$ of an inhomogeneous drive field. The non-uniformity of the couplings $J_i$ is a crucial element of the models under consideration. For perfectly uniform couplings ($J_i = \overline{J}$), the model is integrable and exactly solvable in terms of the macroscopic spin $\boldsymbol{\wS} = \sum_{i} J_i \mathbf{S}_i$. The $S^x S^x$ and $S^y S^y$ terms in~\eqref{eq:SeparableModelDelta} describe spin-exchange interactions between pairs of atoms, mediated by virtual cavity photons, while the $S^z S^z$ terms describe state-dependent ac Stark shifts. The normalization of $H$, which is not important for the dynamical properties, ensures that the high-temperature specific heat and free energy have a proper thermodynamic limit (see Supplementary Material Section~\ref{appendix:Bandwidth}).

The dynamical phases generated by this non-local spin model, shown in Fig.~\ref{fig:PhaseDiagram}, are accessible via two experimentally tunable parameters. The spin-anisotropy parameter $\Delta$, controlling the relative strength of the spin-exchange and $S^z S^z$ interactions, can be tuned by changing the angle of an applied magnetic field $\mathbf{B}$ (see Fig.~\ref{fig:PhaseDiagram}). In addition, it is possible to control the strength of quantum effects by changing the spin size $S$ on each site. While the choice of internal atomic states provides some flexibility in varying $S$, a larger range of spin sizes can be achieved by varying the number of atoms trapped at each site and letting $S^{\alpha}_i$ represent the collective spin of the sub-ensemble at site $i$. This enables the tuning of quantum effects from semi-classical dynamics at large $S$ all the way down to a spin-$1/2$ system that is dominated by quantum fluctuations. In combination with the possibility of varying the anisotropy $\Delta$, this tunability allows for a thorough exploration of the dynamical phase diagram.

The paper and the presentation of the various regimes shown in Fig.~\ref{fig:PhaseDiagram} are organized as follows. We provide a brief overview of these dynamical phases in Section~\ref{sec:OverviewPhaseDiag} and we emphasize the novel features, which constitute our main results. In Section~\ref{sec:ExpScheme} we describe in detail the proposed experimental  scheme to realize and control the couplings of the Hamiltonian~\eqref{eq:SeparableModelDelta}. We also describe ways of inducing perturbations that go beyond separable interactions. In Section~\ref{sec:IntegrablePoints} we begin the derivation of the main results. We analytically construct the integrals of motion that demonstrate the integrability of the dynamics at the special points $\Delta=0$ and $\Delta=1$. In Section~\ref{sec:SVDmethod} we present a computational method for finding integrals of motion using numerical or experimental data. In Section~\ref{sec:ExactDiag} we deploy this technique and provide numerical evidence for the existence of a novel quantum integrable regime away from the special points $\Delta=0,1$. Specifically, we present an exact diagonalization study of the spin-$1/2$ model, 
showing that the integrable structure persists for anisotropy values $\Delta \neq 0,1$ away from the integrable points, with quasi-bilinear integrals of motion.
In Section~\ref{sec:ClassicalNumerics} we simulate the classical model ($S\rightarrow\infty$) and show that it becomes chaotic, albeit in the presence of slowly decaying modes, away from the special points. In Section~\ref{sec:Experiment2} we discuss experimental limitations and assess the extent to which the various features of the model are accessible in the presence of dissipation. Finally, in Section~\ref{sec:Discussion} we comment on the implications of these results before concluding.

\section{Overview of the phase diagram}
\label{sec:OverviewPhaseDiag}

The best understood part of the dynamical phase diagram in Fig.~\ref{fig:PhaseDiagram} is the line at $\Delta=1$, for all spin sizes $S$, on which the Hamiltonian from Eq.~\ref{eq:SeparableModelDelta} is equivalent to a rational Gaudin model~\cite{Gaudin}. This model is quantum integrable in the mathematical sense of possessing an underlying quantum group structure~\cite{Sklyanin}. In the context of Gaudin-type models, quantum integrability is characterized by the existence of an extensive family of commuting, \emph{bilinear} conserved quantities and there exist analytical expressions for each one. Even though there is no notion of spatial locality, the conserved quantities are ``$2$-local'' in the complexity theory sense~\cite{KitaevBook,Aharonov2007}. By interchanging commutators with Poisson brackets, it follows that the integrable structure persists in the classical limit.

We find that the model is integrable at $\Delta=0$ as well. We obtain analytical expressions for an extensive family of conserved quantities that are also bilinear in spin. As in the case of $\Delta=1$, this integrability holds for any value of $S$, including the classical limit $S\to \infty$. The integrability of the model at $\Delta=0$ is connected to the existence of a non-standard class of Gaudin models~\cite{Balantekin2005, Skrypnyk2005,Schmidt2008,Skrypnyk2009art1,Skrypnyk2009art2,Skrypnyk2009art3,Lukyanenko2016,Claeys2019}.

However, the most surprising part of the phase diagram occurs away from these integrable points, i.e. in the regions $\{\Delta<0\}$, $ \{0 < \Delta <1\}$, and  $\{\Delta>1\}$. There, we find a novel integrable structure that is markedly different from the type of integrability found at the two integrable points, $\Delta = 0$ and $\Delta = 1$. First, unlike the latter, integrability for $\Delta \neq 0,\, 1$ appears to depend crucially on the spin size $S$. We show strong evidence that the model is integrable for a spin-$1/2$ system, while it is chaotic with a finite Lyapunov exponent $\lambda_{L}$ in the classical limit ($S\to \infty$). Nevertheless, in this latter limit, we also find that there exist modes that relax only on time scales much larger than $\lambda_{L}^{-1}$. We conjecture that this is a consequence of the ``quasi-integrable'' nature~\cite{Kurchan19quasi} of the classical model. The putative transition from quantum integrable to (semiclassical) chaotic dynamics, schematically shown in Fig.~\ref{fig:PhaseDiagram}, can be probed experimentally.

Second, the integrals of motion (IOM) of the $S=1/2$ model at $\Delta \neq 0, \, 1$ are not bilinear (or 2-local), but may instead be termed $\emph{quasi}$-bilinear. We present compelling numerical evidence that each IOM has appreciable support in the space of bilinear spin operators that does \emph{not} depend on the system size $N$. The fact that the integrals of motion persist while developing tails of multi-spin terms on top of the dominant two-spin contribution is reminiscent of the quasi-local integrals of motion that characterize Many-Body Localized phases~\cite{Vosk2013,Serbyn2013LIOM,Huse2014,Chandran2014,Ros2015}.

\section{Proposed Experimental scheme}
\label{sec:ExpScheme}

As advertised, the full phase diagram of Fig.~\ref{fig:PhaseDiagram}a can be accessed in experiments with atomic ensembles in single-mode optical cavities. In such experiments, each spin is encoded in internal states of an individual atom. The cavity generically couples to a weighted collective spin
\begin{equation}\label{eq:wS}
\vc{\wS} \equiv \sum_i \xi_i \mathbf{S}_i,
\end{equation}
where each weight $\xi_i$ is set by the amplitudes of the cavity mode and drive field at the position of the $i^\mathrm{th}$ atom.  Experiments to date have realized either Ising interactions~\cite{leroux2010implementation,Hosten2016science,Braverman2019,Landini2018,Kroeze2018} $H \propto \wS_z^2$ or spin-exchange interactions~\cite{norcia2018cavity,davis2019photon} $H \propto \wS_+\wS_-$, in the latter case directly imaging the spatial dependence of the weights $\xi_i$ and the resulting spin dynamics~\cite{davis2019photon}. We now show how to extend the approach of Ref.~\onlinecite{davis2019photon} to realize generic XXZ models of the form
\begin{equation}\label{eq:Hxxz}
H \propto \wS_x \wS_x + \wS_y \wS_y + \Delta \wS_z\wS_z,
\end{equation}
where the anisotropy $\Delta$ is tuned by the angle of a magnetic field. An alternative approach to engineering Heisenberg models has been proposed in Ref.~\onlinecite{Mivehvar2019}.

The experimental setup proposed here is shown in Fig.~\ref{fig:PhaseDiagram}b. We consider spins encoded in Zeeman states of atoms whose positions in the cavity are fixed by a deep optical lattice. A magnetic field $\mathbf{B} = B\hat{z}$, which defines the quantization axis for the spins, is oriented at an angle $\theta$ to the longitudinal axis $\hat{c}$ of the optical cavity. Driving the atoms with a control field, incident either through the cavity or from the side, allows pairs of atoms to interact by scattering photons via the cavity. The interaction strengths are governed by the spatially dependent Rabi frequency $\Omega_i$ of the control field and vacuum Rabi frequency $2g_i$ of the cavity, where $i$ denotes the value for the $i^\mathrm{th}$ atom.

For large detuning between the atomic and cavity resonances, the atom-cavity interaction takes the form of a Faraday effect in which each atom couples to the Stokes vector $\mathbf{I}_i$, representing the local polarization and intensity of light. This effect is described by a Hamiltonian
\begin{equation}
    H_I = 2 \chi \sum_i \left(\mathbf{I}_i\cdot\hat{c}\right) \left(\mathbf{S}_i \cdot \hat{c}\right),
    \label{eq:ExpIntHamiltonian}
\end{equation}
where $\chi$ is the vector ac Stark shift of a maximally coupled atom and the component of the Stokes vector along the cavity is $\mathbf{I}_i\cdot\hat{c}\ =  ( \adj{A}_{+,i} A_{+,i} - \adj{A}_{-,i} A_{-,i} )/2$. The field operators
\begin{equation}
A_{\pm,i} = \left(\frac{\Omega_{i} e^{-i \delta t}}{2\sqrt{2}} + g_i a_\pm\right)/g
\end{equation}
include the quantum field $a_\pm$ of the cavity for $\sigma_\pm$-polarized modes, weighted by the local amplitude $g_i$ of the cavity mode, and displaced by a classical drive field with local Rabi frequency $\Omega_i$.  The normalization is set by the vacuum Rabi frequency $2g$ of a maximally-coupled atom.  We assume that the drive field has horizontal polarization $\hat{x} = \hat{z}\times\hat{c}$ and is detuned by $\delta$ from the cavity resonance.

In the limit where the drive field is weak and far detuned, we can obtain an effective Hamiltonian for the spin dynamics by adiabatically eliminating the photon modes. To this end, we first expand $H_I$ to lowest order in the operators $a_\pm$ to obtain
\begin{equation}
H_I \approx \frac{i}{2} \chi \left( \xi_i^* v e^{i\delta t} - \xi_i \adj{v} e^{-i \delta t} \right)\left(\mathbf{S}_i \cdot \hat{c}\right),
\label{eq:ExpIntLowestOrder}
\end{equation}
where $v = \left(a_+ - a_-\right)/\sqrt{2}$ represents the vertically polarized cavity mode, and we have introduced the weights
\begin{equation}
    \xi_i = \frac{\Omega_i g_i^*}{g^2}.
    \label{eq:xi_experimental}
\end{equation}
These weights determine the collective spin $\wS$ defined in Eq.~\ref{eq:wS}, which couples to the cavity mode. Then, for $\avg{\adj{v} v}\ll1$ and for large detuning $\delta \gg \kappa, \omega_Z$ compared to the cavity linewidth $\kappa$ and Zeeman splitting $\omega_Z$, we find that the effective spin Hamiltonian is
\begin{equation}\label{eq:exp_H}
H = \frac{\chi^2}{4 \delta}\left[ \cos^2 \theta \mathcal{F}^z \mathcal{F}^z  + \frac{1}{2}\sin^2 \theta \left(\mathcal{F}^x \mathcal{F}^x   +\mathcal{F}^y \mathcal{F}^y\right) \right],
\end{equation}
as detailed in the Supplementary Material Section~\ref{appendix:ExperimentalH}. We see that Eq.~\ref{eq:exp_H} matches the Hamiltonian~\eqref{eq:SeparableModelDelta} with couplings $J_i = \chi \xi_i S^{1/2} N^{1/4} \sin \theta / 2 \sqrt{2 \delta}$ and anisotropy $\Delta = 2 \cot^2 \theta$. Note that arbitrary control over the set of weights $\xi_i$ can be obtained by designing the spatial dependence of the control field.

In addition to the coherent dynamics generated by $H$ from Eq.~\ref{eq:exp_H}, the cavity-mediated interactions are subject to dissipation due to photon loss from the cavity mirrors and atomic free-space scattering. Formally, these processes can be described by a family of Lindblad operators acting within a quantum master equation (see the Supplementary Material Section~\ref{appendix:ExperimentalDissipation}). The key parameter governing the interaction-to-decay ratio is the single-atom cooperativity $\eta = 4 g^2 / \kappa \Gamma$, where $\Gamma$ is the atomic excited-state linewidth. Moreover, we find that the interaction-to-decay ratio is collectively enhanced, scaling as $S\sqrt{N\eta}$ for a system of $N$ sub-ensembles consisting of $S$ atoms each.

After we discuss the various properties and measurable signatures of chaotic and integrable dynamics in~\eqref{eq:SeparableModelDelta}, we shall return to quantifying the effects of dissipation in Section~\ref{sec:Experiment2}. In particular, we will estimate the atom number and cooperativity $\eta$ requisite for observing these signatures in the experimental setup.

\section{Integrability at $\Delta=0$ and $\Delta=1$}
\label{sec:IntegrablePoints}

In this section, we demonstrate the quantum integrability of the Hamiltonian~\eqref{eq:SeparableModelDelta} along the two lines at $\Delta=0$ and $\Delta=1$ in the dynamical phase diagram (Fig.~\ref{fig:PhaseDiagram}). To place our discussion in context for the non-specialist reader, we begin by recalling some key features of integrable many-body systems. Broadly speaking~\footnote{For a comprehensive discussion of the subtleties involved in achieving a generally valid definition of quantum integrability, see Ref.~\onlinecite{Caux}.}\label{fn}, such systems are characterized by an extensive number of local conservation laws that give rise to exotic transport and thermalization properties.  Important examples of quantum integrable systems include the Lieb-Liniger Bose gas and the spin-$1/2$ Heisenberg chain.

To illustrate the main ideas, consider a one-dimensional, local, quantum Hamiltonian $H=\sum_{n=1}^N h_n$, on $N$ lattice sites. For this type of model, integrability means the existence of $N-1$ independent, local charges,
\begin{equation}
Q^{(n)} = \sum_{i=1}^N q^{(n)}_i, \quad n=2,\ldots,N,
\end{equation}
that commute with each other and with the Hamiltonian, namely
\begin{equation}
\label{intccr}
[Q^{(m)},Q^{(n)}] = 0, \quad [Q^{(m)},H] = 0.
\end{equation}
The existence of extensively many local conservation laws can be regarded as a strong constraint on the dynamics of such systems, and leads to unusual physical effects such as non-dissipative heat transport\cite{zotos} and equilibration to non-thermal steady-states\cite{Rigol,Barthel}.

In contrast with more standard integrable systems, the Gaudin-type models that arise in the present work are somewhat unusual, since they exhibit non-local couplings and are therefore essentially zero-dimensional. To construct these models, one starts from a set of $N$ operators,
\begin{equation}
\label{GeneralSpinH}
G^{(i)} = \sum_{j=1}^N \sum_{\alpha=1}^3 w^\alpha_{ij} S^\alpha_i S^\alpha_j, \quad i=1,2,\ldots, N,
\end{equation}
that are linear combinations of spin bilinears, with real coefficients $w^\alpha_{ij} \in \mathbb{R}$, and satisfy the defining commutation relations:
\begin{equation}
\label{commrel}
[G^{(i)},G^{(j)}] = 0.
\end{equation}
The physical Hamiltonian and the independent conserved charges are then given by linear combinations of the $G^{(i)}$, of the form
\begin{align}
H = \sum_{i=1}^N a^{(0)}_i G^{(i)}, \quad
Q^{(n)} = \sum_{i=1}^N a^{(n)}_i G^{(i)}, \quad n=2,\dots,N,
\end{align}
where the coefficients $a^{(n)}_i \in \mathbb{R}$ are elements of a non-singular $N$-by-$N$ matrix. Note that by the commutation relations~\eqref{commrel}, the Hamiltonian $H$ and its associated charges $Q^{(n)}$ automatically satisfy the commutation relations~\eqref{intccr} required for integrability. Although these operators are not local, they are sums of spin bilinears and can therefore be regarded as ``2-local'' in the complexity theory sense.

We now show that the Hamiltonian Eq. ~\eqref{eq:SeparableModelDelta} defines a Gaudin-type integrable model for $\Delta=0$ and $\Delta=1$ and all values of spin $S$. Specifically, we will demonstrate that along these lines in the dynamical phase diagram Fig.~\ref{fig:PhaseDiagram}, there exist $N-1$ independent, conserved and mutually commuting spin bilinears. The Hamiltonian at $\Delta=1$ is related to the rational Gaudin model~\cite{Gaudin}, which is well-known to be quantum integrable in the mathematically rigorous sense of possessing an underlying quantum group structure~\cite{Sklyanin}. Meanwhile, the Hamiltonian at $\Delta=0$ lies in a less well-known class of ``non-skew'' Gaudin models, which arise from Gaudin's equations upon relaxing the constraint of antisymmetry under interchange of site indices~\cite{Balantekin2005, Skrypnyk2005,Schmidt2008,Skrypnyk2009art1,Skrypnyk2009art2,Skrypnyk2009art3,Lukyanenko2016,Claeys2019}.

It will be helpful to review the problem first studied by Gaudin~\cite{Gaudin}: under what circumstances do a set of spin bilinears, as in Eq. \eqref{GeneralSpinH}, define a mutually commuting set, with $[G^{(i)},G^{(j)}]=0$? If the couplings $w_{ij}^{\alpha} \in \mathbb{R}$ are taken to be antisymmetric under interchange of indices, with $w_{ij}^{\alpha}+w_{ji}^{\alpha}=0$, then the $G^{(i)}$ mutually commute if and only if the \emph{Gaudin equations}
\begin{equation}
\label{eq:GaudinEq}
w_{ij}^{\alpha}w^{\gamma}_{jk}+w^{\beta}_{ji}w^{\gamma}_{ik}-w^{\alpha}_{ik}w^{\beta}_{jk} = 0,
\end{equation}
hold for all pairwise distinct $\{i,j,k\}$ and $\{\alpha, \beta, \gamma\}$. The isotropic solution $w^\alpha_{ij} = J_iJ_j/(J_i-J_j)$ defines the \emph{rational Gaudin Hamiltonians}
\begin{equation}
G^{(i)}(\vec{J}\;) =  \sum_{j \neq i}\frac{J_iJ_j}{J_i-J_j} \mathbf{S}_i \cdot \mathbf{S}_j.
\end{equation}
The all-to-all spin model from Eq.~\ref{eq:SeparableModelDelta} at $\Delta=1$ is simply a linear combination of rational Gaudin Hamiltonians and Casimirs, to wit
\begin{equation}
H = \sum_{i,j=1}^N J_i J_j \mathbf{S}_i \cdot \mathbf{S}_j = \sum_{i=1}^N 2J_i G^{(i)}(\vec{J}) + J_i^2 \mathbf{S}_i \cdot \mathbf{S}_i.
\label{eq:Delta1ConservedQuantities}
\end{equation}
By rotational symmetry, $H$ conserves the total spin $\mathbf{S}_{\mathrm{tot}} = \sum_i \mathbf{S}_i$, and the linear span of the $G^{(i)}(\vec{J}\;)$ includes the squared spin $\mathbf{S}_{\mathrm{tot}}\cdot\mathbf{S}_{\mathrm{tot}} = \sum_{i,j} \mathbf{S}_i\cdot\mathbf{S}_j$. The mathematical structure of traditional Gaudin models has been studied in depth~\cite{Sklyanin,Feigin1994}.

Let us now consider relaxing the constraint of antisymmetric couplings. Then Gaudin's equations~\eqref{eq:GaudinEq} must be augmented by two equations constraining ``on-site'' couplings, which read
\begin{align}
\nonumber (w_{ij}^\beta w_{ji}^\gamma - w^{\beta}_{ji}w^\gamma_{ij}) + 2w^{\alpha}_{ji}(w^{\gamma}_{ii} - w^{\beta}_{ii}) = 0, \\
\label{eq:GaudinEq2}
(w_{ij}^\alpha w_{ji}^\gamma - w^{\alpha}_{ji}w^\gamma_{ij}) + 2w^{\beta}_{ji}(w^{\gamma}_{ii} - w^{\alpha}_{ii}) = 0.
\end{align}
The model from Eq.~\ref{eq:SeparableModelDelta} at $\Delta=0$ arises from an ``non-skew XXZ'' solution $w^1_{ij} = w^{2}_{ij} = J_iJ_j/(J_i^2-J_j^2), \, w^3_{ij} = J_j^2/(J_i^2-J_j^2)$ to the usual Gaudin equation~\eqref{eq:GaudinEq}, augmented by onsite terms $w^{1}_{ii} = w^{2}_{ii} = 1/2, \, w^{3}_{ii} = 0$, which solve Eq.~\ref{eq:GaudinEq2}. The corresponding Gaudin Hamiltonians read
\begin{align}
\nonumber \widetilde{G}^{(i)}(\vec{J}\;) = \sum_{j\neq i} \frac{J_iJ_j}{J_i^2-J_j^2} &\left(S_i^xS_j^x + S_i^yS_j^y \right) + \frac{J_j^2}{J_i^2 - J_j^2}S_i^z S_j^z \\
+& \frac{1}{2}\left(S_i^xS_i^x + S_i^yS_i^y \right).
\label{eq:Delta0ConservedQuantities}
\end{align}
By the Gaudin equations Eq.~\ref{eq:GaudinEq} and Eq.~\ref{eq:GaudinEq2}, these mutually commute and the Hamiltonian~\eqref{eq:SeparableModelDelta} at $\Delta=0$ can be expressed as
\begin{equation}
H = \sum_{i,j=1}^N J_i J_j (S_i^x S_j^x + S_i^y S_j^y) = \sum_{i=1}^N 2J_i^2 \widetilde{G}^{(i)}(\vec{J}\;).
\label{eq:ModelDelta0}
\end{equation}
At spin-$1/2$, this coincides with the Hamiltonian obtained in Ref.~\onlinecite{Lukyanenko2016} or the ``Wishart-SYK'' model~\cite{Iyoda2018}, and can consequently be derived as a special case of the integrable spin-$1/2$ Hamiltonians considered in the recent work Ref.~\onlinecite{Claeys2019}. The integrability of~\eqref{eq:ModelDelta0} for arbitrary spin $S$ was first discussed in Refs.~\onlinecite{Skrypnyk2009art1,Skrypnyk2009art2,Skrypnyk2009art3} (see also the references therein). We conclude that there is an integrable line in the phase diagram of the model~\eqref{eq:SeparableModelDelta} at $\Delta = 0$. By rotational symmetry about the $z$-axis, this Hamiltonian conserves $S^z_{\mathrm{tot}} = \sum_i S^z_i$ and $(S^z_{\mathrm{tot}})^2$ lies in the linear span of the $\widetilde{G}^{(i)}(\vec{J}\;)$. 
Finally, we note that upon replacing commutators with Poisson brackets in the derivation of the Gaudin equations, the integrable structure identified for $\Delta=0$ and $\Delta=1$ remains unaltered in the classical limit ($S\rightarrow\infty$) of the Hamiltonian.

\section{Extracting integrals of motion from numerical or experimental data}
\label{sec:SVDmethod}

Having characterized the integrable structure for $\Delta=0$ and $\Delta=1$, it is natural to ask whether the integrability of~\eqref{eq:SeparableModelDelta} extends to other, more generic, values of the anisotropy: can we find similar extensive sets of commuting bilinear conserved charges for $\Delta \neq 0, \, 1$? To tackle this question in the absence of analytical tools, such as those used in the previous section, we develop a numerical method that enables the systematic search for bilinear (2-local) integrals of motion (IOM). We emphasize that this novel technique can be applied to either numerical or experimental data.

Let us first define a set of 2-local operators $\{\hat{O}_a\}$:
\begin{equation} \hat{O}_a \equiv \frac{3}{S(S+1)} \hat{S}^\alpha_i \hat{S}^\alpha_j, 
\label{eq:SpinBilinears}
\end{equation} 
where $i>j$ and the index $a$ is a shorthand notation for $(i,j,\alpha)$. We note that this family of $3 N(N-1)/2$ operators defines an orthonormal set with respect to the infinite-temperature inner product:
\begin{align}
 \frac{1}{\mathcal{D}} \mathrm{Tr}[ \hat{O}_a^{\dagger} \hat{O}_b ] = \delta_{ab}, \label{eq:orthonormality} 
\end{align} 
where $\mathcal{D} \equiv \mathrm{Tr}[\mathbf{1}] = (2S+1)^{N}$ is the dimension of the Hilbert space.

Now suppose that we can measure, experimentally or numerically, the time evolution of the expectation value $\langle \hat{O}_a(t)\rangle \equiv  \bra{\Phi} \hat{O}_a(t) \ket{\Phi}$, where $\ket{\Phi}$ is a random initial state (i.e. far from any energy eigenstate). A bilinear integral of motion $\hat{I}$ is a special linear combination of the $\hat{O}_a$ that remains constant in time, to wit
\be
\langle \hat{I}\rangle=\langle \hat{I}(t)\rangle \equiv \sum_a u_a \langle \hat{O}_a(t)\rangle = \sum_a u_a \overline{\langle \hat{O}_a\rangle}.
\label{eq:iom}
\ee
Here and below, the overline denotes a time average, such as 
$\overline{\langle \hat{O}_a\rangle} \equiv  \int_0^T \frac{dt}T \langle \hat{O}_a (t)\rangle$ over a time interval $[0,T]$. It is useful to recast the above equation in terms of the following time series matrix:
\begin{equation}
    M_{a,t}\equiv \sqrt{\frac{1}{T}}\left(\langle\hat{O}_a(t)\rangle - \overline{\langle \hat{O}_a\rangle}\right) \,. \label{eq:defM}
\end{equation}
Note that $M_{a,t}$ is a rectangular matrix with $3N(N-1)/2$ rows and a continuum of columns indexed by $t \in [0,T]$, where $T J^{2} \gg 1$. In practice, the time axis is discretized such that the number of columns in $M$ is much larger than the number of rows. We immediately see that, by Eq.~\ref{eq:iom}, a 2-local IOM corresponds to a left zero mode of $M$, i.e. $\sum_a u_a M_{a, t} = 0$ for any $t$.

Thus, to find bilinear IOMs, we want to search for zero modes of $M$. More generally, we can consider the singular value decomposition (SVD) of $M$, or equivalently, the spectrum of the real Hermitian matrix 
\begin{align}
L_{a,b} & \equiv M M^{\dagger} = \int_0^T \frac{d t}{T} M_{a,t} M_{b,t}  \\
& = \sum_{l=1}^{3N(N-1)/2} \sigma_l^2 u_{a,l} u_{b,l} \,.
\end{align}
In the second line, $\sigma_l \ge 0$ are the corresponding singular values of $M$ and $\sigma_l^2$ are the eigenvalues of $L$; $\vec{u}_l$ are the left singular vectors of $M$ and eigenvectors of $L$. Equivalently, $(u_{a,l})_{a,l = 1}^{3N(N-1)/2}$ is a real orthogonal matrix, defining a family of operators
\begin{equation} \hat{Q}_l \equiv \sum_{a=1}^{3N(N-1)/2} u_{a,l} \hat{O}_a \,,\, l = 1, \dots, 3N(N-1)/2 \,, \label{eq:Ql}\end{equation}
which are also orthonormal:
\begin{equation}
    \frac{1}{\mathcal{D}} \mathrm{Tr}[\hat{Q}_l^{\dagger} \hat{Q}_k] = \delta_{lk} \,.
\end{equation}
As mentioned above, $Q_l$ is an integral of motion if and only if $\sigma_l = 0$. Furthermore, for small $\sigma_l>0$, we consider $Q_l$ to be approximately conserved and call it a ``slow mode.'' The rationale for this terminology comes from the identity
\begin{equation}\label{eq:SVD_variance} 
\overline{\left< Q_l (t) \right>^2} - \overline{\left< Q_l (t) \right>}^2 = \sigma_l^2,
\end{equation}
which means that the singular value $\sigma_l$ is the standard deviation of the expectation value of $Q_l$ over the time interval $[0,T]$. A small $\sigma_l$ entails that $\langle Q_l (t) \rangle$ exhibits small fluctuations around its time-average value.

To summarize, we propose the following procedure: compute the time series matrix $M$, perform an SVD decomposition on $M$, analyze its singular values, and identify the possible IOMs and slow modes. In the next two sections, we use this method to characterize the behavior of the model along the $S=1/2$ and $S\rightarrow\infty$ lines in the phase diagram of Fig.~\ref{fig:PhaseDiagram}, for anisotropies $\Delta\neq 0,\, 1$. In Section~\ref{sec:ExactDiag}, we numerically simulate the time evolution for the quantum spin-1/2 model and we further characterize the resulting slow modes by measuring their temporal auto-correlation functions. In Section~\ref{sec:ClassicalNumerics}, we simulate the dynamics of the model~\eqref{eq:SeparableModelDelta} describing classical spin degrees of freedom and, upon slightly modifying the above method, we extract the behavior of the auto-correlation functions directly from the singular values.

\section{Integrability$^*$ for $S=\frac{1}{2}$}
\label{sec:ExactDiag}

\begin{center}
\begin{figure}
\includegraphics[width=\columnwidth]{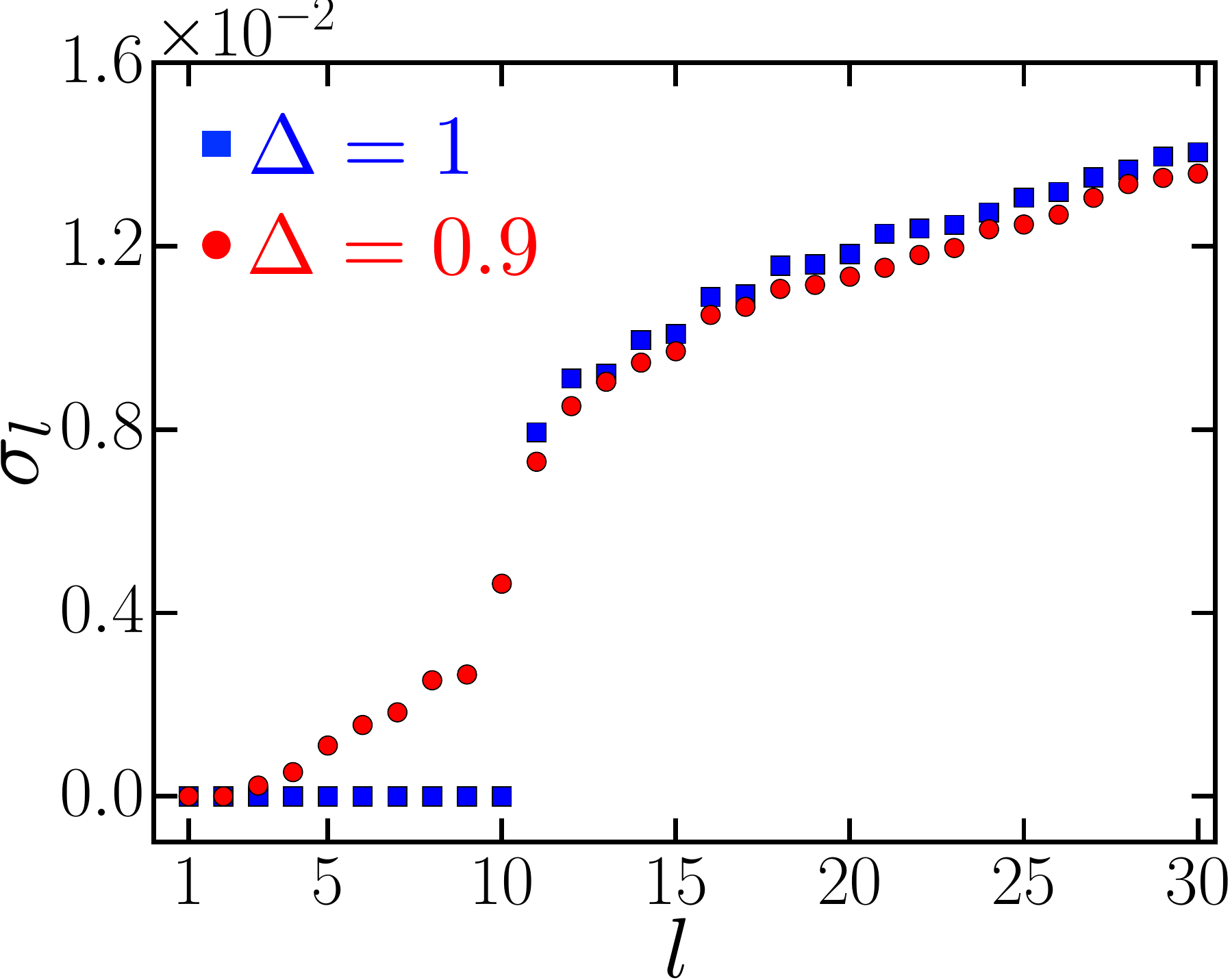}
\caption{Scatter plot of the smallest 30 of the $3N(N-1)/2$ singular values $\sigma_l$ at $N=9$ for $\Delta=1$ (blue squares) and $\Delta = 0.9$ (red circles) in a fixed disorder realization of $\{J_i\}$. At $\Delta = 1$, we see $N+1$ zeros corresponding to the $N+1$ conserved charges that can be written as a sum over bilinear operators; these zeros are separated from the rest of the singular values by a ``spectral'' gap. At $\Delta = 0.9$, we see two zeros corresponding to the conservation of $H$ and $\left(S_{\mathrm{tot}}^{z}\right)^2$. We also see the lift-off of $N-2$ singular values corresponding to the previously conserved bilinear charges at $\Delta=1$. Note that they, too, are separated from the rest by a ``spectral'' gap.}
\label{fig:SingularValues}
\end{figure}
\end{center}

\subsection{Identifying integrals of motion}
\label{subsec:QuantumIOMs}

We now focus on the spin-$1/2$ system with up to $N=14$ sites and implement the technique proposed above. We initialize the system in a random product state~\footnote{We initialize each spin $i$ in a randomly chosen (with equal probability $p=1/6$) eigenstate of either $\hat{S}_i^x$, $\hat{S}_i^y$ or $\hat{S}_i^z$.} $\ket{\Phi}$ and numerically compute the time evolution of the wavefunction with the Hamiltonian~\eqref{eq:SeparableModelDelta} via exact diagonalization. The random fields $J_i$ are sampled from the normal distribution $\mathcal{N}(0,J^2)$ and we set $J^2=1$. We have checked that we obtain similar results for other distributions with zero mean and unit variance. We then record the expectation values of all the operators $\hat{O}_a$ defined in Eq.~\ref{eq:SpinBilinears} and construct the time series matrix $M_{a,n}$ (defined in Eq.~\ref{eq:defM}) at each discrete time $t_n=n\delta t$ with $\delta t=1\;J^{-2}$, integer $n$, and up to a maximal time $T=10^3 J^{-2}$.

Fig.~\ref{fig:SingularValues} presents results for the singular values of $M$ obtained for two values of $\Delta$ in a fixed disorder realization. As expected, at $\Delta=1$ we find $N+1$ vanishing singular values, in agreement with the analysis of Section~\ref{sec:IntegrablePoints}. All other singular values lie above a gap of about $0.01$, indicating that there are no other 2-local integrals of motion beyond those identified in Section~\ref{sec:IntegrablePoints}.

The results at $\Delta=0.9$, slightly away from the integrable point, are markedly different. We find only two exactly vanishing singular values corresponding to the space spanned by the two obvious integrals of motion, $H$ and $\left(S^z_{\mathrm{tot}}\right)^2$. This behavior persists on the entire open segment $\Delta\in(0,1)$, showing unambiguously that there are no other purely bilinear integrals of motion in this range. Nonetheless, we see that the remaining set of $N-2$ nontrivial IOMs at $\Delta=1$ are transformed, upon moving to the point $\Delta=0.9$, into left singular vectors with non-zero yet small singular values. It stands to reason that these small singular values correspond to operators that exhibit a slow decay because the system is close to the $\Delta=1$ integrable point. We now test this hypothesis by directly examining the decay of these putative ``slow modes.'' 

\begin{center}
\begin{figure}
\includegraphics[width=\columnwidth]{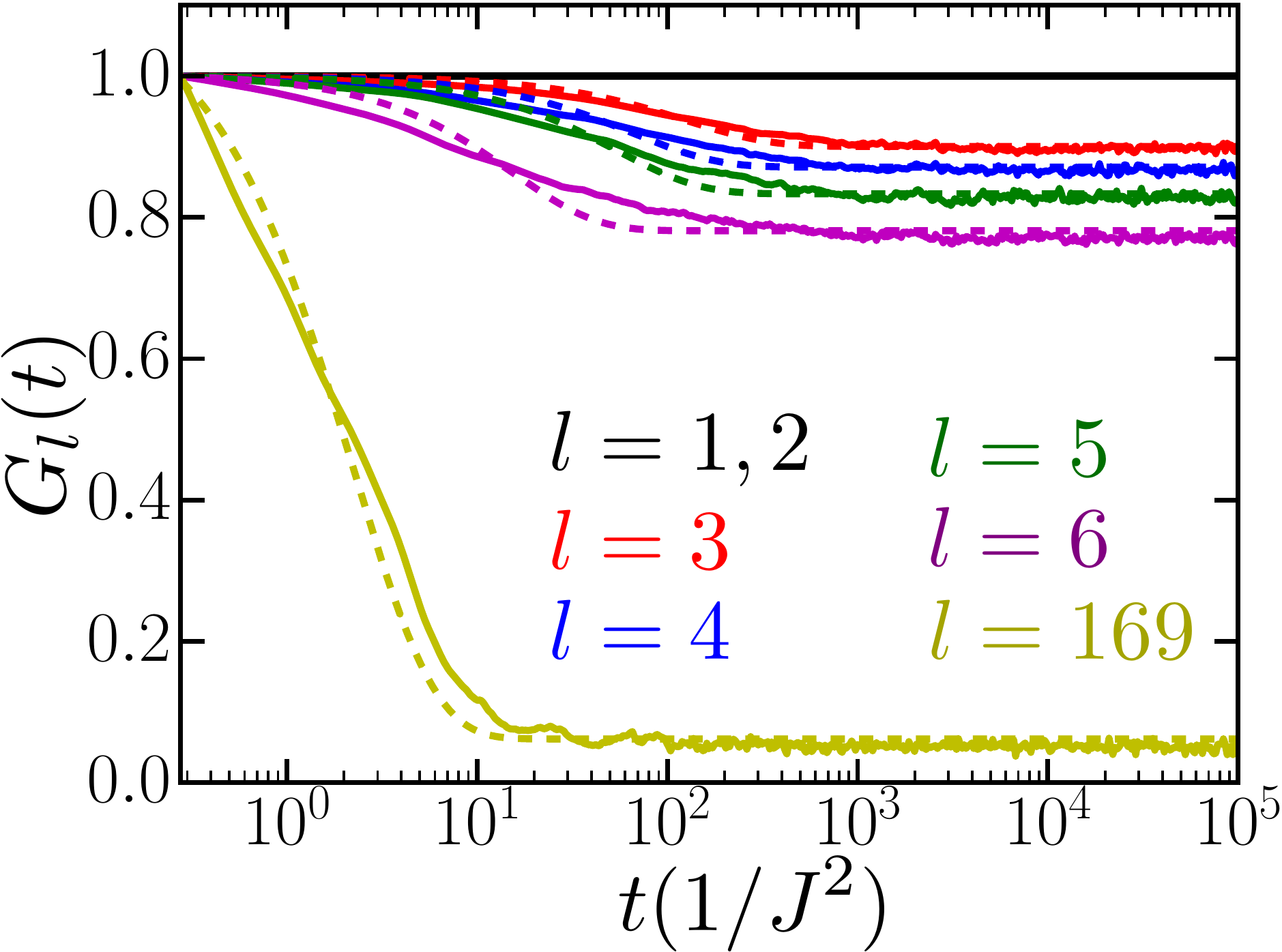}
\caption{Plot of the auto-correlation function $G_l(t)$ in a given disorder realization for $N=13$ spins at $\Delta=0.75$. The solid curves represent the numerically computed $G_l(t)$: the black curve corresponds to either of the two exactly conserved bilinear quantities; the red, blue, green, and magenta curves correspond to the next four modes (arranged by increasing singular value); the yellow curve corresponds to a mode in the middle of the singular value ``spectrum.'' The dashed curves represent fits of the form $\widetilde{G}_l(t) = \zeta_l \exp\left(-t/\tau_l\right) + g_{l}$ through the data.}
\label{fig:ModeFits}
\end{figure}
\end{center}

\subsection{Characterizing the slow operators}
\label{subsec:ModesDecayQuantum}

\begin{center}
\begin{figure}[t]
\includegraphics[width=\columnwidth]{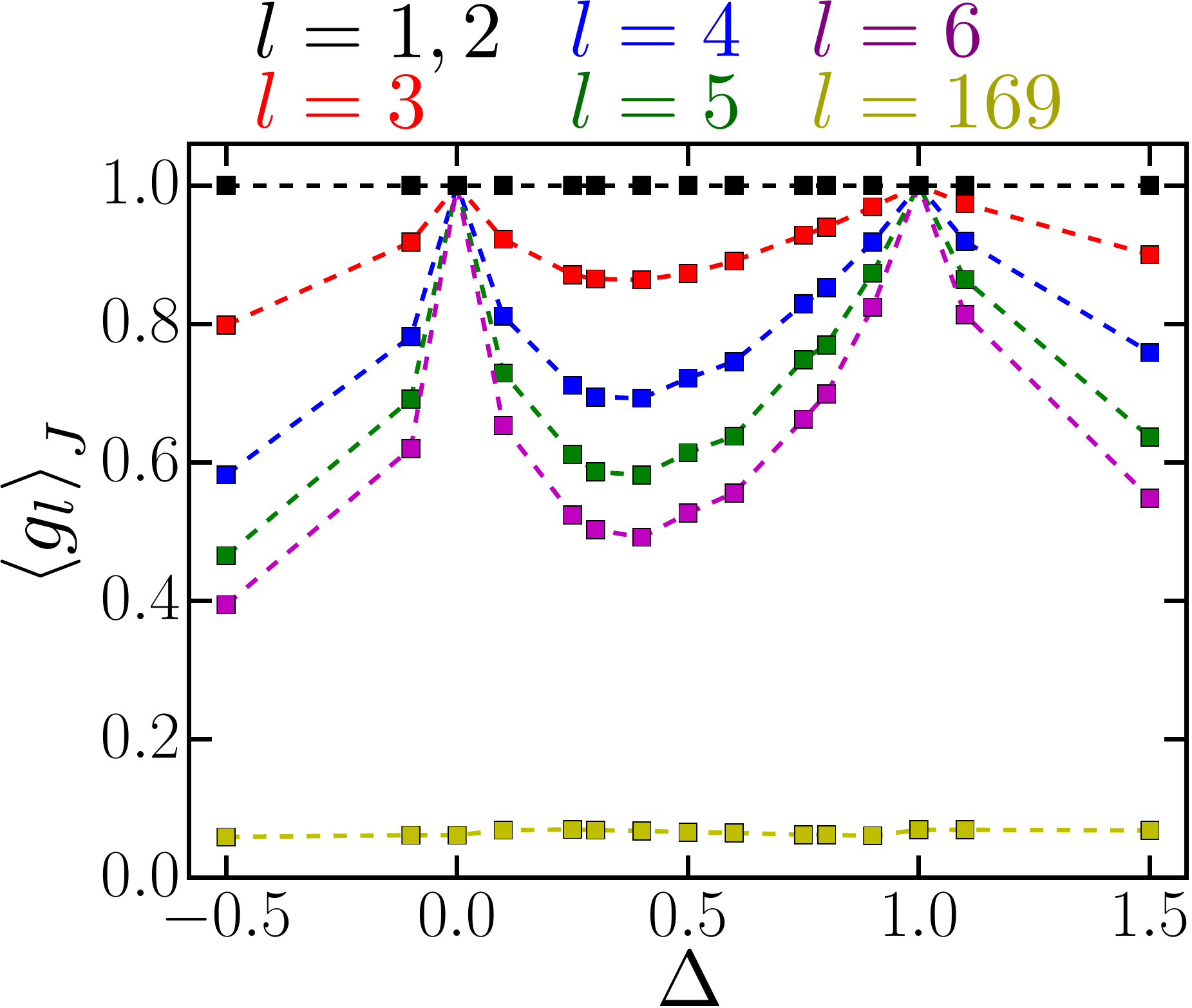}
\caption{Plot of the plateau values $\left< g_{l}\right>_{J} = \left< G_l\left(t\rightarrow \infty\right)\right>_{J}$ as a function of the anisotropy $\Delta$ for $N=11$ spins ($S=1/2$). The brackets $\left<\dots\right>_{J}$ denote an average over $2000$ disorder realizations for the $\{J_i\}$. Different colors correspond to different modes: black corresponds to the two lowest and exactly conserved modes; red, blue, green, and magenta correspond to the next four modes; yellow corresponds to a mode in the middle of the singular value ``spectrum''. We find no strong dependence on the system size $N$: see Fig.~\ref{fig:PlateausDelta05} for a plot of the plateau value $\left< g_l\right>_{J}$ as a function of the system size $N$ for the $l=3$ (red) mode at $\Delta=0.5$.}
\label{fig:Plateaus}
\end{figure}
\end{center}

We have seen that the nontrivial IOMs at the points $\Delta=0$ and $\Delta=1$ transform into a set of $N-2$ ``slow operators,'' indicated by small singular values, away from those two points. Let us examine the dynamics of these presumed slow modes. Their decay can be studied by numerically computing the auto-correlation functions
\begin{equation}
G_l(t) = \frac{1}{\mathcal{D}} \Tr\left[\hat{Q}_l(t) \hat{Q}_l(0)\right],
\label{eq:InfTempAutoCorr}
\end{equation}
where the normalization $\mathcal{D} = \left(2S+1\right)^{N}$ ensures that $G_l(0) = 1$. For conserved modes, we expect the auto-correlation function to remain fixed at $G_l(t) = 1$ for all time. For generic non-conserved operators, we expect $G_l(t)$ to decay to values near zero as these modes thermalize.

\begin{center}
\begin{figure*}
\includegraphics[width=\textwidth]{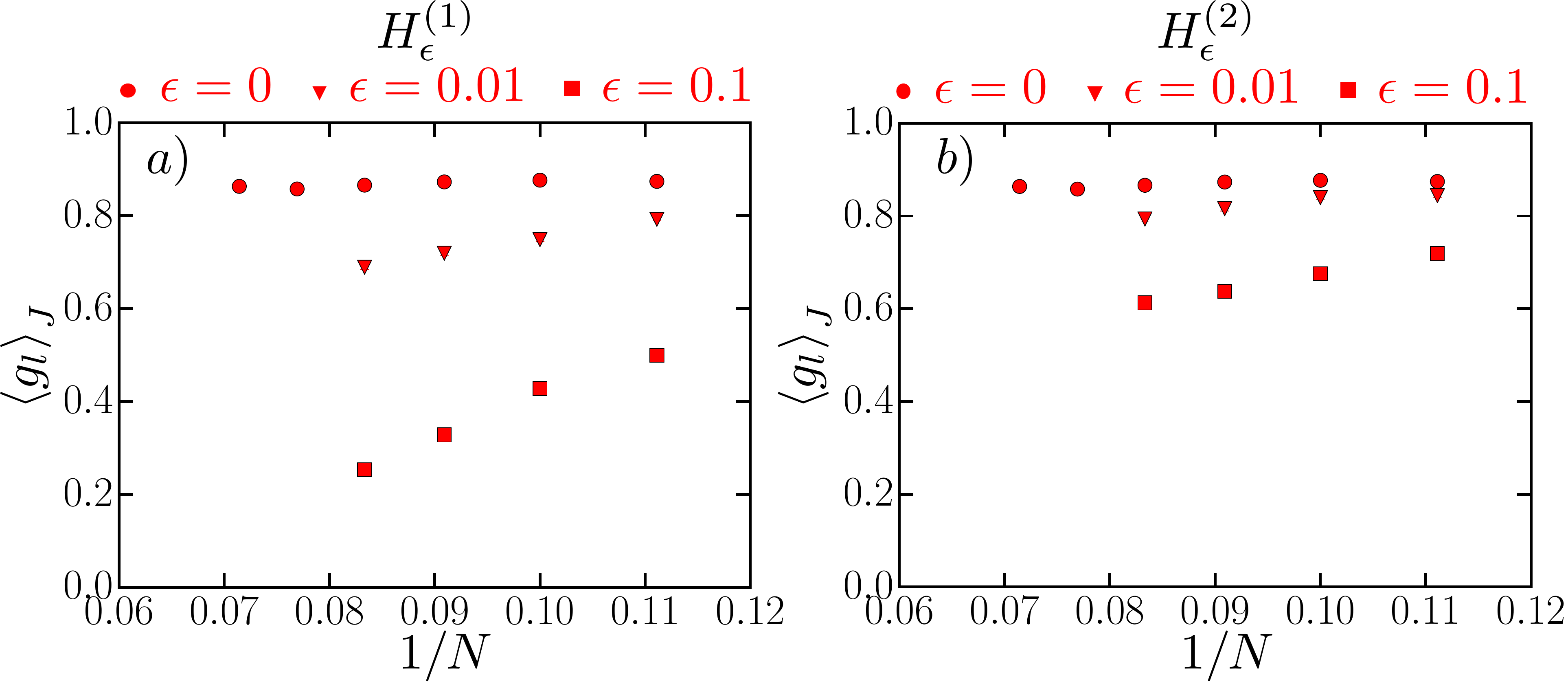}
\caption{Plot of the disorder-averaged plateau values $\left< g_{l}\right>_{J} = \left< G_l\left(t\rightarrow \infty\right)\right>_{J}$ as a function of the system size $N$ for $l=3$ mode, i.e. the lowest mode that is not exactly conserved (corresponding to the red markers in Fig.~\ref{fig:Plateaus}) at $\Delta=0.5$. The different markers correspond to various strengths $\epsilon$ of the perturbations $H^{(1)}_{\epsilon}$ (left panel) and $H^{(2)}_{\epsilon}$ (right panel) from Eq.~\ref{eq:ChaoticPerturbation} and Eq.~\ref{eq:UncorrelatedFields}, respectively: the round markers correspond to the unperturbed Hamiltonian $H$~\eqref{eq:SeparableModelDelta}; the triangular and square markers correspond to $\epsilon=0.01$ and $\epsilon=0.1$, respectively. The error bars related to disorder averaging $\left<\dots\right>_{J}$ are included, but they are smaller than the size of the markers. We see that the plateau value for the unperturbed $H$ is independent of the system size. Converserly, upon adding even a small perturbation $\epsilon \ll 1$, the plateau value decreases with $N$, suggesting that the autocorrelation function $G_l\left(t\rightarrow \infty\right)$ vanishes for a thermodynamic system ($N \rightarrow \infty$).}
\label{fig:PlateausDelta05}
\end{figure*}
\end{center}

An example of the results for a system with $N=13$ sites and $\Delta=0.75$ is shown in Fig.~\ref{fig:ModeFits}. We see that the correlation functions related to the two zero singular values, $G_1(t)$ and $G_2(t)$, are perfectly non-decaying, as they must be. Also as expected, the correlation functions $G_l(t)$ associated with the small non-vanishing singular values ($3\leq l\leq N$) show a slow initial decay. However, the surprise is that, at very long times, these correlation functions saturate to a non-vanishing and rather appreciable value $g_l$. Fig.~\ref{fig:Plateaus} shows that this phenomenon persists when varying $\Delta$ on the segment $[-0.5,1.5]$. Moreover, we find no significant size dependence of the saturation value $g_l$, as shown in Fig.~\ref{fig:PlateausDelta05}a. We have also checked that the large plateau values are not due to the overlap between the slow modes $\hat{Q}_l$ with higher powers of the known conservation laws $\hat{H}$ and $\hat{S}^z_{\text{tot}}$, such as $\hat{H}^{2},\hat{H}^{3}, \dots $, nor with projectors to energy eigenstates  (for details, see the Supplementary Material~\ref{appendix:OverlapH2}). In contradistinction, the operators corresponding to higher singular values ($l \gg N$) decay to a vanishing, or very small, saturation value (see the Supplementary Material~\ref{appendix:MiddleSVD}).

Altogether, in addition to the obvious bilinear IOMs, $H$ and $(S^z_{\mathrm{tot}})^2$, we find $N-2$ operators whose correlation functions saturate to an appreciable non-vanishing value. This result suggests that the model remains integrable even away from the Gaudin-like points $\Delta=0$ and $\Delta=1$: the bilinear integrals of motion are transformed into quasi-bilinear ones, which retain appreciable support in the space of 2-local operators. Based on the results shown in Fig.~\ref{fig:Plateaus}, we argue that this holds everywhere away from the integrable points, namely in the regions $\{\Delta<0\}$, $ \{0 < \Delta <1\}$, and  $\{\Delta>1\}$. In general, we can write the new integrals of motion as bilinear operators dressed by a sum over higher, $2n$-local terms:
\begin{equation}
    \hat{I}_{l} = Z_{l} \hat{Q}_l + \sum_{n>1} \sum_{i_1,\dots,i_{2n}} \sum_{\alpha_1,\dots,\alpha_{2n}} K_{i_1\dots i_{2n}}^{\alpha_1\dots\alpha_{2n}} \hat{S}_{i_1}^{\alpha_1}\hat{S}_{i_2}^{\alpha_2}\dots\hat{S}_{i_{2n}}^{\alpha_{2n}},
    \label{eq:qiom}
\end{equation}
where $Z_l$ is the weight of the integral of motion $I_l$ on 2-local operators. The saturation value of the auto-correlation function of $\hat{Q}_l$ that we plot in Fig.~\ref{fig:Plateaus} is, essentially, $g_l\sim|Z_l|^2$. It would be interesting to further characterize how the coefficients $K_{i_1\dots i_{2n}}^{\alpha_1\dots\alpha_{2n}}$, which encode the overlap of the IOMs with the different $2n$-body spin operators, decay with increasing $n$. We leave this for future work.

The structure of the integrals of motion~\eqref{eq:qiom} is, in some ways, reminiscent of the Local Integrals of Motion (LIOM) in the Many-Body Localized (MBL) state~\cite{Vosk2013,Serbyn2013LIOM,Nandkishore2015}. The latter is characterized by quasi-local integrals of motion $\tau_i^z$ that are adiabatically connected to the microscopic degrees of freedom $\sigma^z_i$. As in our case, the LIOMs are dressed versions of the microscopic bits with weight on higher $n$-body operators decaying exponentially with $n$. There are, however, crucial differences from MBL. The integrals of motion in our case are not local, but rather extensive sums of bi-local operators. Hence, the IOMs of the all-to-all spin model do not facilitate a  direct-product partition of the Hilbert space into single qubit spaces. Additionally, the integrability we observe does not depend on strong disorder---in fact, we found that its signatures are more pronounced as the couplings becomes more uniform, namely as $\mathrm{std}(J_i) \lesssim \overline{J_i}$.

Lastly, we also find signatures of integrability in the spectrum of $H$: the level statistics are almost perfectly Poissonian at $\Delta=0,1$ and close to Poisson (although not exactly) at intermediate $\Delta$ (see Supplementary Material section~\ref{appendix:LevelStatistics}). Nonetheless, for $0<\Delta<1$ we find many level crossings and the violation of the Wigner-von Neumann non-crossing rule represents further evidence of integrability despite the fact that there seems to be some degree of correlation between the energy levels~\cite{Schliemann2010,Yuzbashyan2009,Yuzbashyan2016}.

\subsection{Perturbing away from integrability$^*$}
\label{subsec:SpoilingIntegrability}

After establishing the existence of a novel integrable structure for the spin-1/2 model, characterized by quasi-2-local IOMs, it is natural to investigate its robustness to perturbations away from the class of models~\eqref{eq:SeparableModelDelta} with separable disorder. This question is relevant from a theoretical point of view, but also from a practical, experimental perspective.

A natural perturbation to test in this context is one that adds a non-separable, SY-like, contribution to the interaction. Specifically, we add the term
\begin{equation}\label{eq:ChaoticPerturbation}
H^{(1)}_{\epsilon} = \frac{\epsilon}{2S\sqrt{N}}\sum_{i,j=1}^{N} V_{ij}\left[S_{i}^{x}S_{j}^{x} + S_{i}^{y}S_{j}^{y} + \Delta S_{i}^{z}S_{j}^{z}\right],
\end{equation}
where the elements $V_{ij}$ are also sampled from a normal distribution $\mathcal{N}(0,1)$.  

We explicitly check that at $\epsilon>0$ and $\Delta=1$ for $H+H_{\epsilon}^{(1)}$ there are only 4 zero singular values corresponding to exactly conserved and linearly-independent 2-local quantities: the Hamiltonian, $S_{\mathrm{tot}}^2$, $\left(S_{\mathrm{tot}}^{x}\right)^2$, and $\left(S_{\mathrm{tot}}^{y}\right)^2$. At intermediate $0<\Delta<1$, there are only two vanishing singular values corresponding to $H+H_{\epsilon}^{(1)}$ and $\left(S_{\mathrm{tot}}^{z}\right)^2$. Second, we verify that the lowest bilinear modes that are not exactly conserved (i.e. either the $l=3$ one at $0<\Delta<1$ or the $l=5$ one at $\Delta=1$) decay to smaller plateau values which \emph{decrease} as we increase the system size $N$, as shown in Fig.~\ref{fig:PlateausDelta05}a. This suggests that a perturbation $H_{\epsilon}^{(1)}$, even at $\epsilon \ll 1$, can spoil the integrability for a large system $N \gg 1$.

Another type of perturbation that arises naturally in the experimental set-up, due to the driving field, is represented by random stray magnetic fields along the $z$-axis:
\begin{equation}\label{eq:UncorrelatedFields}
    H^{(2)}_{\epsilon} = \epsilon \sum_{i=1}^{N} h_i S_{i}^z,
\end{equation}
where the fields $h_i$ are also sampled from $\mathcal{N}(0,1)$. Note that $H + H^{(2)}_{\epsilon}$ has a single zero singular value corresponding to $\left(S_{\mathrm{tot}}^z\right)^2$ for all $\Delta$; this is due to the fact that the full Hamiltonian is no longer \emph{purely} bilinear and that $H^{(2)}_{\epsilon}$ breaks the $\mathrm{SU(2)}$ symmetry at $\Delta=1$. Aside from this effect, the behavior upon perturbing with $H^{(2)}_{\epsilon}$ is similar to that obtained by perturbing with $H^{(1)}_{\epsilon}$, as shown in Fig.~\ref{fig:PlateausDelta05}b.

Last, we consider the effect of adding the perturbation
\begin{equation}\label{eq:FlipFlopperturbation}
    H^{(3)} = \frac{1}{S\sqrt{N}} \sum_i J_i^2 S_{i}^z.
\end{equation}
This term appears in the model
\begin{equation}
    \widetilde{H} = \frac{1}{S\sqrt{N}}\sum_{ij} J_iJ_j \left(S_i^{+}S_{j}^{-} + \Delta S_{i}^z S_{j}^z\right),
    \label{eq:FlipFlopH}
\end{equation}
which is similar to Eq.~\ref{eq:SeparableModelDelta}, but differs from it by the term $H^{(3)}$, arising due to the commutator $\left[S_i^{+},S_{i}^{-}\right]$. As noted in Ref.~\onlinecite{Marino2018}, the model Eq.~\eqref{eq:FlipFlopH} is also experimentally accessible in a system of cold atoms interacting with cavity photons. It is clear that the perturbation $H^{(3)}$, having a $1/\sqrt{N}$ normalization, is sub-extensive and will not matter in the thermodynamic limit. Moreover, we find that it does \emph{not} qualitatively affect the integrability of our quantum model even for the small systems considered in ED (see the Supplementary Material~\ref{appendix:Commutator} for the numerical results).

In sum, our numerical analysis of the response to perturbations indicates that the novel integrability of the spin-1/2 model~\eqref{eq:SeparableModelDelta} is not particularly robust to non-separable interactions or stray magnetic fields. Nevertheless, in a finite-size system and at finite times (see Section~\ref{sec:Experiment2} for more details), there are signatures of proximate integrability, as shown by the finite saturation values in Fig.~\ref{fig:PlateausDelta05}.

To recapitulate our study of the dynamical phase diagram Fig.~\ref{fig:PhaseDiagram} thus far, we have found that the system is integrable along three lines: at $\Delta=0,1$ for any value of the spin size $S$ (characterized by bilinear IOMs), and at $S=1/2$ for any $\Delta \neq 0,1$ (characterized by quasi-bilinear IOMs). The remaining line in the phase boundary of Fig.~\ref{fig:PhaseDiagram} corresponds to the classical, $S\rightarrow\infty$, limit of the model~\eqref{eq:SeparableModelDelta}, which we now discuss.

\begin{center}
\begin{figure*}
\includegraphics[width=\textwidth]{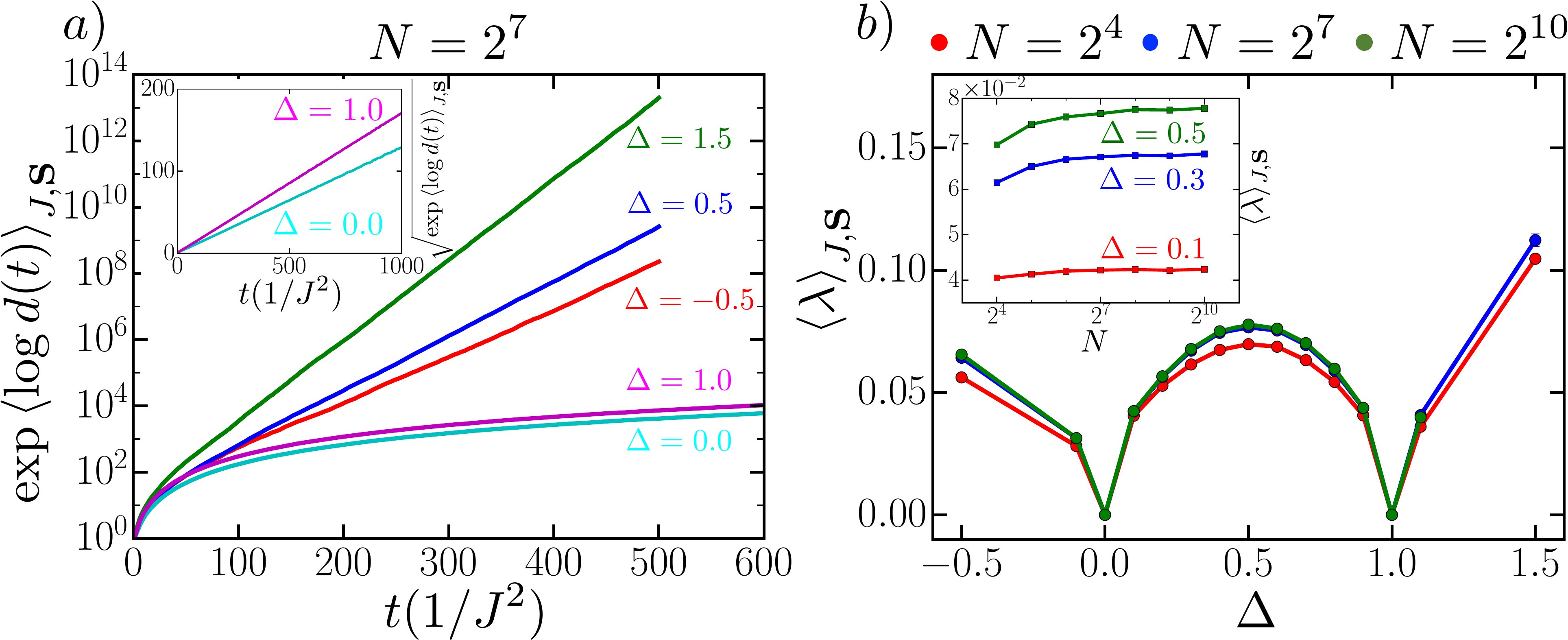}
\caption{a) The sensitivity $d(t)$ from Eq.~\ref{eq:Sensitivity} geometrically averaged over $10^4$ disorder realizations $\{J_i\}$ and initial states $\{\mathbf{S}(0)\}$ as a function of time for a system of $N=128$ classical spins. After non-universal dynamics at early times, we find an exponential growth at later times for $\Delta = -0.5$ (red curve), $\Delta=0.5$ (blue curve), and $\Delta = 1.5$ (green curve). (inset) The square root of the same quantity, namely $\sqrt{\exp\left(\left<\log d(t)\right>_{J,\mathbf{S}}\right)}$, at the two integrable points, $\Delta = 0.0$ (cyan curve) and $\Delta=1.0$ (magenta curve): we obtain an almost perfect straight line, which indicates that $d(t) \sim t^2$, as expected for an integrable system.\\
b) Lyapunov exponent $\left<\lambda\right>_{J,\mathbf{S}}$ averaged over $10^5$ disorder realizations and initial states as a function of the anisotropy $\Delta$ for different systems consisting of $N=16$ (red circles), $N=128$ (blue circles), and $N=1024$ (green circles) classical spins. (inset) Disorder-averaged Lyapunov exponent $\left<\lambda\right>_{J,\mathbf{S}}$ as a function of the system size $N$ for $\Delta = 0.1$ (red squares), $\Delta = 0.3$ (blue squares), and $\Delta = 0.5$ (green squares).}
\label{fig:ClassicalChaos}
\end{figure*}
\end{center}

\section{Chaos for $S \to \infty$}
\label{sec:ClassicalNumerics}

Since Gaudin-type integrability at $\Delta=0, \, 1$ persists for all values of the spin size $S$, it is natural to ask whether the integrability* structure at $S=\frac12$, presented in the previous section, also survives for larger values of $S$. Although it is numerically challenging to extend the exact diagonalization study of the previous section to intermediate $S$, the limit $S \to \infty$ leads to classical equations of motion that are amenable to numerical simulation.

These simulations allow us to analyze another boundary in the phase diagram, namely the $S \to \infty$ line, where we find chaotic dynamics with a finite Lyapunov exponent, as explained in Section ~\ref{sec:cl_chaos}.
The presence of chaos in the infinite-$S$ limit clearly implies that the $S=\frac12$ integrability$^*$ does not extend to all $S$, unlike the Gaudin-type integrability at $\Delta=0$ and $1$. Remnants of an integrability$^*$ structure can nevertheless be revealed by applying the SVD analysis of Section~\ref{sec:SVDmethod} to the classical dynamics, which we do in Section~\ref{subsec:ClassicalIOMs}.
This technique reveals the presence of a large number of slow modes, which are known to occur classically in ``quasi-integrable'' systems, i.e. chaotic systems in the vicinity of integrable points.
We characterize these slow modes in Section~\ref{subsec:ModesDecayClassical}.

\subsection{Classical chaos}
\label{sec:cl_chaos}

In the infinite-$S$ limit, the model~\eqref{eq:SeparableModelDelta} behaves as a classical system of coupled spin degrees of freedom $S_i^{\alpha}$ on the unit sphere, whose Hamiltonian dynamics can be written in terms of Poisson brackets:
\begin{equation}
    \frac{d S_i^{\alpha}}{dt} = \{S_i^{\alpha},H \},
    \label{eq:ClassicalEOMs}
\end{equation}
where
\begin{equation}
    H = \frac{S}{2\sqrt{N}}\sum_{ij}J_iJ_j \left(S_{i}^{x}S_{j}^{x} + S_{i}^{y}S_{j}^{y} + \Delta S_{i}^{z}S_{j}^{z}\right).
    \label{eq:ClassicalHamiltonian}
\end{equation}

For our numerical investigation, we sample the random fields $J_i$ from the uniform distribution $\left[-J,J\right]$ and set $J=1$ (we choose a bounded distribution to avoid large $J_i$'s that could cause numerical instabilities). The classical spin variables $S_i^{\alpha}$ obey
\begin{equation}
    \{S_i^{\alpha}, S_j^{\beta} \} = \frac{1}{S}\delta_{ij} \varepsilon^{\alpha \beta \gamma} S_i^{\gamma}.
    \label{eq:PoissonAlgebra}
\end{equation}
We shall probe the infinite-temperature dynamics of this classical system by direct numerical simulation. 

In order to study chaos, we use the standard tangent space method~\cite{Benettin1980} to study the divergence of classical trajectories and measure the leading Lyapunov exponent. Let $\mathbf{S}(t) = \left(\mathbf{S}_1(t),\dots,\mathbf{S}_N(t)\right)$ denote the $3N$-dimensional vector describing the directions of all the spins at time $t$. We initialize the system in a random infinite-temperature state $\mathbf{S}(0)$, within which each spin points in a random direction, uniformly distributed on the unit sphere $S_2$. We also keep track of the trajectory of the deviation vector $\delta\mathbf{S}(t)$, which lives in the tangent space of $S_{2}\times \dots \times S_2$ at the point $\mathbf{S}(t)$; we further set $\delta\mathbf{S}(0)$ such that $\delta\mathbf{S}_{i}(0) \perp \mathbf{S}_i(0)$ for all spins and $\norm{\delta\mathbf{S}(0)}^2 = \sum_{i,\alpha} \left(\delta S_i^{\alpha}\right)^2 = 1$.

If we define the local effective field $\mathbf{F}_i = \left(F_i^x,F_i^y,F_i^z\right) = \frac{1}{S}\left(-\frac{\partial H}{\partial S_i^x},-\frac{\partial H}{\partial S_i^y},-\frac{\partial H}{\partial S_i^z}\right)$, we see that the Hamilton equations of motion~\eqref{eq:ClassicalEOMs} can be written as
\begin{equation}
    \frac{d \mathbf{S}_i}{dt} = \mathbf{S}_i \times \mathbf{F}_i.
    \label{eq:SEOMS}
\end{equation}
For our model~\eqref{eq:ClassicalHamiltonian} we have $F_i^{x,y} = \frac{J_i}{\sqrt{N}}\sum_{j}J_j S_j^{x,y}$ and $F_{i}^z = \Delta\frac{J_i}{\sqrt{N}}\sum_{j}J_j S_j^{z}$. 

We immediately see that the variational equations of motion for the deviation vector $\delta \mathbf{S}$ can be written as
\begin{equation}
    \frac{d (\delta \mathbf{S}_i)}{dt} = \delta \mathbf{S}_i \times \mathbf{F}_i + \mathbf{S}_i \times \delta \mathbf{F}_i,
    \label{eq:TangentSEOMS}
\end{equation}
where $\delta F_i^{x,y} = \frac{J_i}{\sqrt{N}}\sum_{j}J_j \delta S_j^{x,y}$ and $\delta F_{i}^z = \Delta\frac{J_i}{\sqrt{N}}\sum_{j}J_j \delta S_j^{z}$. 

We numerically integrate the coupled differential equations~\eqref{eq:SEOMS} and~\eqref{eq:TangentSEOMS} to find the trajectory $(\mathbf{S}(t),\delta\mathbf{S}(t))$ in the tangent bundle up until a time $T = 500 J^{-2}$ in increments $\delta t = 1\;J^{-2}$. We then compute the sensitivity, defined as $d(t) = \norm{\delta \mathbf{S}(t)}^2$, or in full,
\begin{equation}
    d(t) = \sum_{i=1}^{N} \sum_{\alpha} \left[\delta S_{i}^{\alpha}(t)\right]^2.
    \label{eq:Sensitivity}
\end{equation}
Note that $d(0)=1$, since we have normalized the initial deviation vector. For an integrable system we expect $d(t)$ to exhibit a power-law dependence on time; the flow on invariant tori specified by the $N$ conservation laws is linear in time and, since we have defined the sensitivity as $\norm{\delta \mathbf{S}(t)}^2$, we expect $d(t) \sim t^2$. In a chaotic system $d(t)$ should increase exponentially with $t$. In Fig.~\ref{fig:ClassicalChaos}a, we average over disorder realizations $\{J_i\}$ and initial states $\{\mathbf{S}(0)\}$ to find $\exp\left(\left< \log d(t)\right>_{J,\mathbf{S}}\right)$. We find that the classical system exhibits chaotic dynamics and an exponential divergence of trajectories in the regions $\{\Delta<0\}$, $\{0<\Delta<1\}$, and $\{\Delta>1\}$. We also find integrable dynamics and a power law divergence of trajectories at the special points $\Delta=0,1$.

Moreover, using the multiplicative ergodic theorem, we can define the maximal Lyapunov exponent~\cite{Benettin1980} as
\begin{equation}
    \lambda = \lim_{t\rightarrow \infty}\frac{2}{t} \log \frac{\norm{\delta \mathbf{S}(t)}}{\norm{\delta \mathbf{S}(0)}}.
\end{equation}
Using the normalization $\norm{\delta \mathbf{S}(0)} = 1$ and our definition of the sensitivity from Eq.~\ref{eq:Sensitivity}, we see that
\begin{equation}
    \lambda = \lim_{t\rightarrow\infty}\frac{1}{t} \log d(t).
\end{equation}

In practice, we compute the Lyapunov exponent by fitting a line $\lambda t + b$ through the late time behavior of $\log d(t)$, as discussed in Ref.~\onlinecite{Goldhirsch1987}. In Fig.~\ref{fig:ClassicalChaos}b we plot the Lyapunov exponent $\left< \lambda\right>_{J,\mathbf{S}}$, averaged over disorder realizations $\{J_i\}$ and initial states $\{\mathbf{S}(0)\}$, as a function of the anisotropy $\Delta$ and find that the system exhibits the most chaotic behavior (largest Lyapunov exponent) at $\Delta = 1.5$. Second, we find that $\left< \lambda\right>_{J,\mathbf{S}}$ tends to a finite value for large system sizes $N$, as shown in the inset of Fig.~\ref{fig:ClassicalChaos}b.

\begin{center}
\begin{figure*}
\includegraphics[width=\columnwidth]{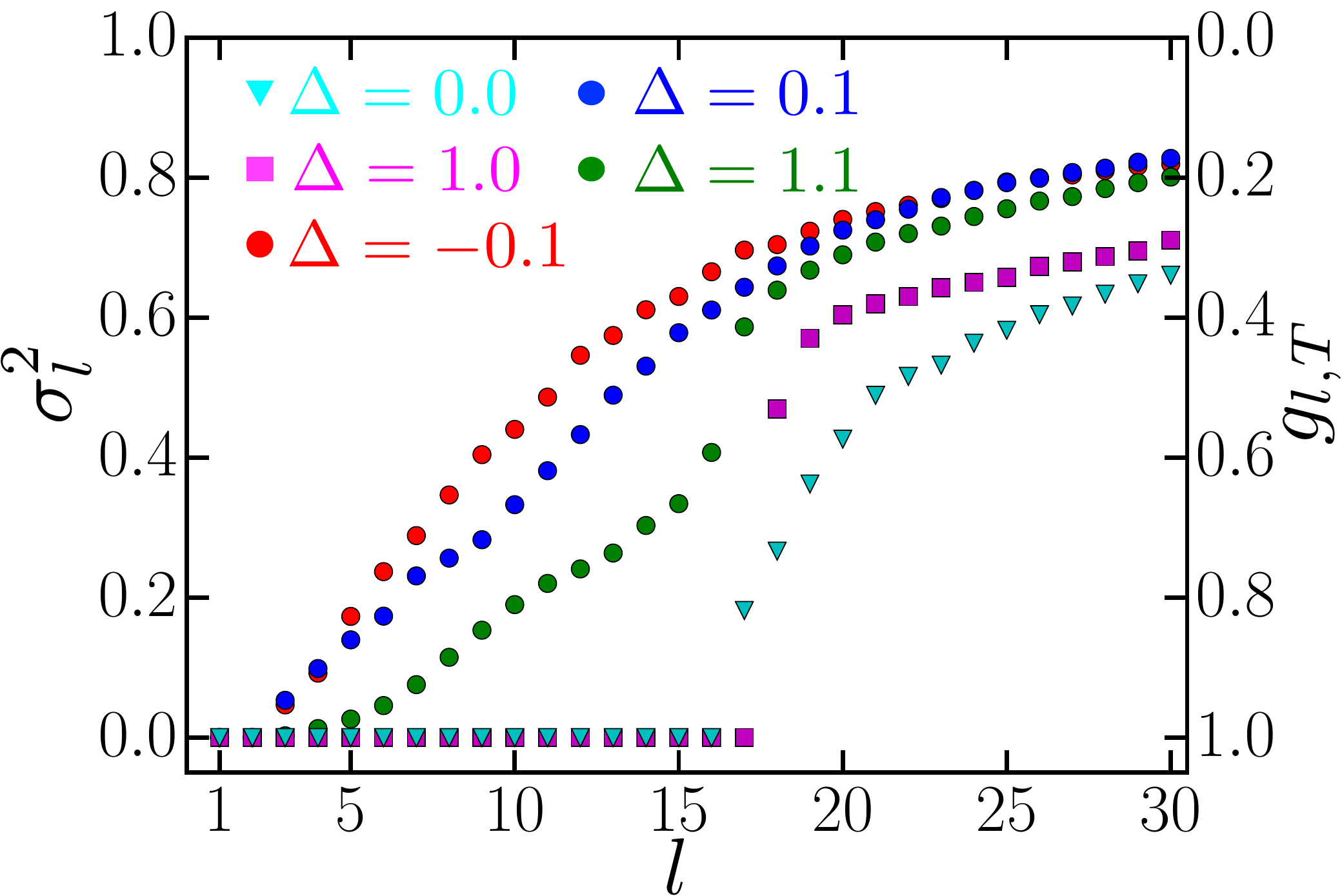}
\caption{Scatter plot of the smallest 30 squared singular values $\sigma_l^{2}$ and the plateau value $g_{l,T}$ from Eq.~\ref{eq:ClassicalPlateauExpression} for a classical model~\eqref{eq:ClassicalHamiltonian} of $N=16$ spins at different values of the anisotropy $\Delta$. We average $\sigma_l^2$ over $\mathcal{N} = 10^3$ random initial conditions $\{S_0\}$ and $10$ disorder realizations for the fields $\{J_i\}$ and we set $T= 8192 J^{-2
}$. At the integrable points $\Delta=0$ (cyan triangles) and $\Delta = 1$ (magenta squares) we see $N$ and $N+1$ zero singular values, respectively, corresponding to the conserved quantities that can be written as a sum over bilinear operators~\eqref{eq:ClassicalBasis}. These are separated from the rest of the singular values by a ``spectral'' gap. At $\Delta=-0.1$ (red circles), $\Delta =0.1$ (blue circles), and $\Delta=1.1$ (green circles) we see two precisely zero singular values corresponding to the conservation of $H$ and $\left(S_{\mathrm{tot}}^{z}\right)^2$, along with the lift-off of the other $N-2$ singular values. For $\Delta = 1.1$, the first $N$ singular values are also separated by a spectral gap from the rest. And, although for $\Delta = -0.1$ and $\Delta=0.1$ the spectral gap is not visible, one can still see a rounded ``cusp'' occuring around $l = N$---this suggests that there still exist slow modes $Q_l$ for $l=3,\dots,N$.}
\label{fig:SVDClassical}
\end{figure*}
\end{center}

\subsection{SVD analysis}
\label{subsec:ClassicalIOMs}

Although the presence of chaos in the classical dynamics excludes proper integrability in the infinite-$S$ limit, it does not rule out the possibility of ``quasi-integrability,'' whereby some operators have very slow decay. We investigate this possibility by applying the SVD analysis of Section~\ref{sec:SVDmethod} to the classical dynamics. This allows us to determine the number of exactly conserved quantities, corresponding to zero singular values, but also to look for slow modes, corresponding to small but finite singular values.

As expected, we find an extensive number of conserved quantities at $\Delta = 0,1$ and only 2 exactly conserved quantities, corresponding to the Hamiltonian $H$ and $\left(S^z_{\mathrm{tot}}\right)^2$, for all other values of the anisotropy $\Delta$. 
This intermediate regime, however, exhibits a large number of slow modes, which will be discussed in the next section.

Since we are now working with a classical system, a few important distinctions ought to be made from our earlier, quantum analysis. First, we consider a slightly enlarged collection of bilinear operators:
\begin{equation}
    O_a = 
    \begin{cases}
    3 S_i^{\alpha} S_j^{\alpha} & a = (i,j,\alpha) \,,\, i < j \\
    c_{1} \left( S_{i}^{x}S_{i}^{x} - 
   S_{i}^{y}  S_{i}^{y} \right)  &  a = (i,i,1) \,,   \\
    c_{2} \left( 3 S_{i}^{z}S_{i}^{z}  - 1\right),&  a = (i,i,2) \,,
    \end{cases}
    \label{eq:ClassicalBasis}
\end{equation}
where $c_{1} = \sqrt{15}/2 $ and $c_{2} = \sqrt{5}/2$. As before, $a = (i,j,\alpha)$ is a composite index. In the classical case, we also include bilinears with $i=j$ (which would be trivial in the spin-$1/2$ case). Note that there are only two independent such bilinears for each $i$, and the spherical harmonics (with spin $1$) provide an orthonormal basis. Indeed, it can be checked that the bilinears $O_a$ defined in Eq.~\ref{eq:ClassicalBasis} satisfy the orthonormality relation
\begin{equation}
 \left< O_a  O_b \right>_{\mathbf{S}} \equiv \int \prod_i \frac{D \mathbf{S}_i}{4\pi}  O_a  O_b = \delta_{ab},
\end{equation}
where $\left< [\dots] \right>_{\mathbf{S}}$ denotes an average over the infinite temperature ensemble, while the integral $\int D \mathbf{S}_i$ is over the unit sphere.

Second, while a single initial state is sufficient in the quantum SVD analysis, we have to consider an ensemble of initial states in the classical setting. This is because a single classical trajectory cannot visit the whole phase space due to energy conservation (a linear superposition of configurations does not exist classically). Here, we take the infinite-temperature ensemble, namely we sample $S_0 = \{\mathbf{S}_1(0), \dots, \mathbf{S}_N(0)\}$ as independent random points on the unit sphere. We then time evolve with~\eqref{eq:ClassicalEOMs} for a total time $T$, and measure the expectation value of the bilinears $O_a(t_n, S_0)$ at discrete intervals $t_n = n\delta t \in [0, T]$. Repeating this for a large number $\mathcal{N}$ of initial conditions $\{S_0\}$ in the infinite-temperature ensemble, we construct the following matrix, analogous to the one in Eq.~\ref{eq:defM}:
\begin{equation}
    M_{a,(t,S_0)} =  \left(O_a(t,S_0) - \overline{O_a}(S_0)\right), \label{eq:defM_classical}
\end{equation}
where $\overline{O_a}(S_0) = \int_0^T \frac{dt}{T}  O_a(t, S_0)$ represents the time average over one trajectory. The number of rows indexed by $a$ is, according to Eq.~\ref{eq:ClassicalBasis}, $3N(N-1)/2 + 2N$. The columns are indexed by time $t \in [0,T]$ and initial condition $S_0$---in practice, we discretize the time axis (with the time-step $\delta t = 1 J^{-2}$) and draw a large number ($10^3$) of samples for $S_0$. 

Then the singular value decomposition of $M$ is equivalent to diagonalizing the real Hermitian matrix $L \equiv MM^{\dagger}$, which can be obtained by averaging over the initial conditions $S_0$:
\begin{equation}
    L_{a,b} = \left< \int_0^T \frac{d t}{T} M_{a,(t_n,S_0)} M_{b,(t_n,S_0)} \right>_{\mathbf{S}}.
\end{equation}
Note that the average $\left< \dots\right>_{\mathbf{S}}$ is with respect to $S_0$ in the infinite-temperature thermal ensemble and should not be confused with the quantum expectation values $\left<\dots\right>$ (i.e. without a subscript) used in Sections~\ref{sec:SVDmethod} and~\ref{sec:ExactDiag}. Diagonalizing $L$ allows us to obtain the slow mode operators $Q_l$, together with their corresponding eigenvalues $\sigma_l^2$. Similarly to~\eqref{eq:SVD_variance}, we have
\begin{align}
    \sigma_l^2& = 
    \left< \overline{Q_l(t,S_0)^2} - \overline{Q_l(t,S_0)}^2  \right>_{\mathbf{S}} \,.
    \label{eq:SingularValueFluctuations}
\end{align}
In other words, $\sigma_l^2$ is equal to the variance, averaged over initial conditions $S_0$, of the fluctuations of $Q_l$ along a given trajectory.

The behavior of the singular values $\sigma_l$ (shown in Fig.~\ref{fig:SVDClassical}) is similar, in several ways, to that obtained in Section~\ref{subsec:QuantumIOMs} for the quantum spin-$1/2$ model~\footnote{The main difference between the classical and quantum SVD results is that the classical singular values are considerably larger than the quantum ones (compare Fig.~\ref{fig:SingularValues} and Fig.~\ref{fig:SVDClassical}). This is because the expectation value of $O_a$ is evaluated on a single classical configuration in the former case, while the quantum expectation value is a result of a coherent average.}. At the first integrable point $\Delta=0$, we obtain $N$ zero singular values corresponding to the family of $N$ spin-bilinear conserved quantities $\widetilde{G}^{(i)}$ from Eq.~\ref{eq:Delta0ConservedQuantities}. At the second integrable point $\Delta=1$, we find $N+1$ zero singular values corresponding to the conserved quantities lying in the linear span of the $G^{(i)}$s from Eq.~\ref{eq:Delta1ConservedQuantities}. Lastly, as shown in Fig.~\ref{fig:SVDClassical}, away from these integrable points, i.e. for $\{\Delta<0\},\{0<\Delta<1\}$, and $\{\Delta>1\}$, we find two precisely zero singular values, corresponding to the two exactly conserved spin-bilinear quantities, $H$ and $\left(S_{\mathrm{tot}}^{z}\right)^2$. The small magnitude of the following singular values, for $l=3,4,\dots$, signals the presence of slow modes, which will be studied in the next section.

\subsection{Decay of slow operators}
\label{subsec:ModesDecayClassical}

The SVD analysis of the previous section revealed a  large number of operators with small singular values.
In principle, we could characterize the thermalization (or lack thereof) of these operators $Q_l$ using, in analogy to the quantum case, a two-point correlation function
\begin{align}
    G_l(t)& = \left< Q_l(t) Q_l(0)\right>_{\mathbf{S}}   \label{eq:ClassicalCorrelationFunction},
\end{align}
where $\left<\dots\right>_{\mathbf{S}}$, as before, designates an average over the $\mathcal{N}$ initial conditions $S_0$. As in the quantum case, the operators $Q_l$ are orthonormal such that $G_l(0) = 1$ (as $\mathcal{N} \to\infty$). Yet, the accurate computation of $G_l(t)$ at long times is typically very demanding because it requires
averaging an increasingly complex function in phase space.

Fortunately, in classical systems, the singular value $\sigma_l$ already informs us about the long time plateau value of $G_l(t)$. This can be seen from Eq.~\ref{eq:SingularValueFluctuations}, which implies that
\begin{align}
    \sigma_l^2 =& \left< \int_0^T \frac{dt}T Q_l(t)^2 \right>_{\mathbf{S}} - 
    \left< \int_0^T \int_0^T   \frac{ds}T \frac{dt}T Q_l(t)Q_l(s)  \right>_{\mathbf{S}}  \nonumber \\ 
     =& 1 - \left< \int_0^T \int_0^T   \frac{ds}T \frac{dt}T Q_l(t)Q_l(s)  \right>_{\mathbf{S}} \nonumber  \\
     \sigma_l^2 =& 1 - g_{l,T} \;,\; g_{l,T} \equiv  \int_0^T  G_l(u)\frac{2(T-u)d u}{T^2}     \,. 
      \label{eq:SVD_G}
\end{align}
In the second line, we used the normalization $\left< Q_l(t)^2 \right>_{\mathbf{S}} = 1$; in the third line, we performed a change of variables $u = |t-s|$ (recall that $G_l(u) = \left< Q_l(t) Q_l(t \pm u) \right>_{\mathbf{S}}$ by the invariance of the infinite-temperature ensemble under time evolution). Now, it is not hard to show that $g_{l,T}$ and $G_l$ have the same infinite-time limit if that exists for $G_l$:
$$ \lim_{u\to \infty}G_l(u) = g_l  \; \Rightarrow \; \lim_{T\to \infty} g_{l,T}
=   g_l \,. $$
Thus, $g_{l,T}$ is a finite-time proxy for $g_l$. In the infinite-time limit, the relation~\eqref{eq:SVD_G} becomes 
\begin{equation}
    g_l = 1 - \sigma_l^2\vert_{T\to\infty}.
    \label{eq:ClassicalPlateauExpression}
\end{equation}
Using Eq.~\ref{eq:SVD_G} or~\ref{eq:ClassicalPlateauExpression}, this allows us to infer the plateau values of slow modes from the data of Fig.~\ref{fig:SVDClassical}. Unsurprisingly, the exactly conserved quantities have $g_l=1$. Away from the integrable points at $\Delta=0,1$, we find that the slowest non-conserved modes, corresponding to $l = 3, 4, \dots, $ (the distinction between the slow modes and the rest is less sharp here than in the quantum case, and it is suggested by the rounded cusp around $l=N$ in Fig.~\ref{fig:SVDClassical} ), have a remarkably slow decay: the plateau values $g_{l,T}$ at a finite but large time $T=10^4J^{-2}$ are close unity~\footnote{We have checked that this persists up to $T=10^6J^{-2}$. Nonetheless, we find that these plateaus eventually decay for a finite-size system, albeit after a very long time. We leave the quantitative analysis for future work.}, comparable to their spin-$1/2$ counterparts. 

In a future companion paper, we will demonstrate that any conserved operator in the $S=1/2$ quantum case is approximately conserved in the $S\rightarrow\infty$ classical model as well, up to $1/N$ corrections in large systems. Therefore, the classical model is expected to display some signatures of integrability. In this section, we saw that such signatures cannot be found from the Lyapunov exponent, but only from the relaxation of slow modes. This is intriguing, but a similar phenomenon has previously been observed. Ref.~\onlinecite{Kurchan19quasi} showed that for certain systems near integrability (called ``quasi-integrable'' by the authors), the relaxation time of certain operators can be significantly longer than the finite Lyapunov time $1/\lambda_L$. Given that our classical system is surrounded by integrable lines $\Delta = 0,1$ and (arguably) $S=1/2$ in the $(\Delta,S)$ parameter plane (see Fig.~\ref{fig:PhaseDiagram}), we conjecture that it is also quasi-integrable. From this perspective, the existence of slow modes is compatible with the finite classical Lyapunov exponent found in Section~\ref{sec:cl_chaos}.

\section{Experimental Realities}
\label{sec:Experiment2}

We have provided analytical and numerical evidence for the rich dynamical phase diagram depicted in Fig.~\ref{fig:PhaseDiagram}, including clear signatures of chaotic dynamics at large $S$ and $\Delta \neq 0,1$, along with signatures of integrability at the special points $\Delta = 0,1$ for any $S$. Further, we demonstrated signatures of a novel integrability* phase at $S=1/2$ for $\Delta \neq 0,1$. We now discuss prospects for observing these signatures in the laboratory. First, what should one measure to identify the chaotic and integrable regimes of the phase diagram? Second, given the inevitable presence of dissipation in realistic experiments, what are the requirements on cavity cooperativity to access the relevant time-scales experimentally?

To identify integrals of motion, the SVD method of Section~\ref{sec:SVDmethod} can equivalently be implemented with experimental data. Using state-sensitive imaging of the atomic ensemble~\cite{davis2019photon}, one may immediately extract the bilinear spin correlation functions $\langle O_a(t) \rangle \propto \langle S_i^{\alpha}(t) S_j ^{\alpha} (t) \rangle$ defined in Eq.~\ref{eq:SpinBilinears}. As each image is obtained from a destructive measurement, one must repeat the experiment many times to obtain statistics of the spin bilinears at a fixed time $t$, and then repeat this procedure for many time-points $t$ to obtain the full matrix $M_{a,t}$. With this matrix in hand, one can then directly apply the singular-value decomposition performed above in Section~\ref{sec:SVDmethod}.

A caveat is that measurements of the spin bilinears can be affected by dissipation due to photon loss and atomic free-space scattering.  Photon loss from the cavity mode causes a random walk in the orientation of the weighted collective spin $\boldsymbol{\wS}$ defined in Eq.~\ref{eq:wS}.  This effect is described by Lindblad 
operators
\begin{subequations}
\begin{align}
L_{\pm} &= \sqrt{\gamma/2} \ \wS_\pm,\\
L_{z} &= \sqrt{\Delta\gamma} \ \wS_z,
\end{align}
\end{subequations}
where the decay rate (derived in Sec.~\ref{appendix:ExperimentalDetails}) is given by
\begin{equation}\label{eq:gamma}
\gamma = \frac{J^2}{S \sqrt{N}}\frac{\kappa}{\delta}.
\end{equation}
The collective dissipation can be suppressed by increasing the detuning $\delta$ and compensating with increased drive strength, until limited by free-space scattering.

The effect of free-space scattering is to project or flip individual spins, as described by a set of Lindblad operators
\begin{equation}
L_{n,(m,m')} = \sqrt{C_{m,m'}\Gamma_\mathrm{sc}}\ket{m}\bra{m'}_n,
\end{equation}
where $m$ or $m'$ indicates the spin state of an individual atom indexed by $n$, and $C_{m,m'}$ is an order-unity branching ratio.  At large detuning, the scattering rate scales as
\begin{equation}\label{eq:Gsc}
\Gamma_{\mathrm{sc}} \sim \frac{J^2}{\eta S\sqrt{N}} \frac{\delta}{\kappa}.
\end{equation}
Comparing Eqs.~\ref{eq:gamma} and~\ref{eq:Gsc} shows that the cooperativity $\eta$ will dictate an optimal detuning for minimizing the net effect of the two forms of dissipation, with higher cooperativity enabling increasingly coherent dynamics.

To determine the cooperativity required to observe the signatures of integrability, we first write down explicit equations of motion for the spin bilinears $\langle S_i^{\alpha}(t) S_j^{\alpha}(t) \rangle$ evolving under the influence of pure collective dissipation or pure single-atom decay, respectively (see Supplementary Material Section~\ref{appendix:ExperimentalDetails}).  We find that the spin bilinears decay exponentially at a rate $\Gamma_{\mathrm{sc}}$ due to free-space scattering and at a rate $\gamma$ due to photon loss from the cavity.  Notably, the rate of spin relaxation due to photon loss is \textit{not} superradiantly enhanced, thanks to the counterbalanced effects of the $L_\pm$ Lindblad operators.  Thus, at weak to moderate cooperativity $\eta\lesssim 1$ and large detuning $\delta>\kappa$, free-space scattering dominates and the bilinears decay on a time-scale $\tau J^2 \sim \sqrt{N} \eta S \kappa/\delta$.  For strong coupling $\eta \gg 1$, where free-space scattering is suppressed relative to cavity decay, the total dissipation can be minimized at a detuning $\delta \sim \sqrt{\eta}\kappa$, leading the spin bilinears to decay on a time-scale $\tau J^2 \sim \sqrt{N \eta} S $.

To compare the decay time $\tau$ with the characteristic time-scales for observing the signatures of integrability, we refer to the time dependence of the autocorrelation functions $G_l(t)$ shown in Fig.~\ref{fig:ModeFits}.  To observe the slow modes, a minimum requirement is to evolve the system for a time $t \gtrsim t^{*} \approx 10 \ J^{-2}$, which governs the rapid decay of all non-integrable autocorrelation functions.  This time can be reached even at $S=1/2$ in a strong-coupling cavity $\eta\sim 10$ with a system of $N=10^3$ sites, or with weaker single-atom cooperativity at larger $S$.  To observe the plateaus themselves, we must evolve the system for a significantly longer time, at least $t \approx 10^3 \ J^{-2}$ according to Fig.~\ref{fig:ModeFits}, which places a more stringent requirement $\sqrt{N}\eta S \gtrsim 10^3 \delta/\kappa$.  This regime is challenging to access for $S=1/2$ but readily accessible with large-$S$ subensembles, e.g., at $\eta \gtrsim 1$ with $N = 10^2$ sites each consisting of $S = 10^3$ spin-1 atoms.

Thus, current experiments are well positioned to explore the regime of mesoscopic spin $S$, in between the quantum ($S=1/2$) and classical ($S\rightarrow \infty$) limits.  This will allow for testing the prediction that the plateaus in $G_l(t)$ calculated for spin $S=1/2$, indicating integrability across the full range $\{\Delta<0\}$,$\{0<\Delta<1\}$, and $\{\Delta>1\}$, persist for larger spin $S$ up to $1/N$ corrections (see Sec.~\ref{subsec:ModesDecayClassical}).  Experiments with scalable spin size $S$ may furthermore shed light on the transition from quantum integrability to chaos in the classical limit, as signified by the positive Lyapunov exponent in Fig.~\ref{fig:ClassicalChaos}.

The chaotic dynamics observed in the classical limit $S \rightarrow \infty$ can be studied experimentally via the hallmark of sensitivity to perturbations. Recent theoretical and experimental work has shown that such sensitivity is accessible in quantum systems by measuring out-of-time-order correlators (OTOCs)~\cite{Larkin1969, Shenker2014, Maldacena2016bound, Hosur2016, swingle2016measuring, Garttner2017,Vermersch2018,Lewis-Swan2019}, which quantify the spread of operators in time via the commutator $C(t) = \left \langle [V(t), W(0)]^2 \right \rangle$. The connection to classical chaos is made clear in the semi-classical limit: for operators $V = S_i^z$, $W = S_j^z$, one can show that, to lowest order in a $1/S$ expansion, $C(t) \propto ( \partial S_i^z(t) / \partial \phi_j )^2$ for a small rotation $\phi_j$ at site $j$ about the $z$-axis~\cite{COTLER2018318}.  Thus, semi-classically the OTOC $C(t)$ measures the sensitivity of the coordinate $S_i^z(t)$ to changes in initial conditions $S_j^z(0)$, and may therefore be regarded as a quantum generalization of the classical sensitivity $d(t)$ defined in Section~\ref{sec:cl_chaos}.

One way to access out-of-time-order correlators experimentally is to ``reverse the flow of time'' by dynamically changing the sign of the Hamiltonian~\cite{swingle2016measuring, Garttner2017}. In the cavity-QED system considered here~\cite{davis2019photon}, this sign reversal is achieved by switching the sign of the laser detuning $\delta$ in Eq.~\ref{eq:exp_H}.  The resilience of such time-reversal protocols to experimental imperfections, including dissipation, has been analyzed theoretically in Ref.~\onlinecite{swingle2018resilience}.

To allow for probing chaos in the cavity-QED system proposed here, the rates of collective dephasing and of decoherence via single-atom decay must be small compared with the Lyapunov exponent.  We thus require $\lambda\tau\gg 1$, where $\tau$ is the characteristic decay time defined above.  More specifically, given the Lyapunov exponents $\lambda \leq 0.08J^2$ shown in Fig.~\ref{fig:ClassicalChaos}, and the requirement of observing the system for several Lyapunov ``decades'' to clearly identify exponential growth (Fig.~\ref{fig:ClassicalChaos}a), we would like to evolve the system for times $t\gtrsim 100 \ J^{-2}$, which are readily accessible in the large-$S$ regime that is of interest for approaching the classical limit.

Even in this regime, the light leaking from the cavity produces a continuous weak measurement of the collective spin $\boldsymbol{\wS}$ whose quantum back-action may have consequences for the dynamics.  The interplay of measurement back-action with chaos in open quantum systems, while beyond the scope of the present work, is a subject of active inquiry~\cite{eastman2017tuning,xu2019extreme} and of fundamental importance for elucidating the quantum-to-classical transition~\cite{habib1998decoherence}.  The proposed experimental scheme, including the possibility of tuning the strength and form of coupling to the environment, opens new prospects for exploring this interplay.

\section{Discussion}
\label{sec:Discussion}

We have studied a class of spin models with separable, all-to-all, random interactions and found a complex dynamical phase structure that depends on the spin size $S$ and the anisotropy $\Delta$ along the $z$-axis. We showed that our model at $\Delta=1$ is equivalent to the well-studied rational Gaudin model, and exhibits special integrable dynamics for all values of $S$. We also proved and confirmed numerically that there exists another special point at $\Delta = 0$ where the model is also integrable (in the same sense), regardless of the spin size. Surprisingly, we found compelling numerical evidence that the system at $S=1/2$ is integrable for any anisotropy $\Delta\notin\{0,1\}$. In contrast to the special points $\Delta=0,1$, the integrals of motion at other values of $\Delta$ are \emph{not} purely spin bilinears and develop tails on $2n$-body terms. We leave the detailed characterization of these dressing tails to future work. Lastly, we found that integrability away from $\Delta = 0,1$ is a purely quantum phenomenon: by numerically solving the Hamilton equations of motion for the classical model ($S\rightarrow \infty$), we showed that its dynamics is chaotic with a non-zero Lyapunov exponent and that there exist only two exactly conserved quantities, as opposed to the extensive family of conservation laws characterizing a classically integrable system. However, even in the classical regime we find an extensive number of quasi-conserved charges, whose decay time appears to diverge in the large-$N$ limit. A more thorough study of this regime will be given in future work.

Our analysis opens up several further lines of inquiry. First, since the Hamiltonian~\eqref{eq:SeparableModelDelta} at the special point $\Delta=1$ (and, presumably, at $\Delta=0$ as well~\cite{Balantekin2005,Skrypnyk2005,Skrypnyk2009art1,Skrypnyk2009art2,Skrypnyk2009art3}) possesses a quantum group structure, does the integrable$^*$ phase exhibit any algebraic structure? Is it possible to construct explicitly the dressed conserved quantities in terms of the model couplings?

Second, we have seen that even though the level statistics of the spin-1/2 system deviates from Wigner-Dyson statistics, exhibiting many level crossings (this holds also for spin-1, as shown in the Supplementary Material), its classical counterpart is chaotic with a finite Lyapunov exponent. We note that this does not contradict the Berry-Tabor conjecture~\cite{BerryTabor,bohigas}, which applies to the semiclassical, large-$S$, regime. In fact, the same phenomenon is known to occur in integrable quantum spin chains, such as the anisotropic Heisenberg model (or XXZ chain): its Hamiltonian 
$\sum_j \left[S^x_j S^x_{j+1} + S^y_j S^y_{j+1} + \Delta S^z_j S^z_{j+1}\right]$ is quantum integrable only for spin-1/2, and it is classically chaotic; its \textit{integrable} higher-spin extensions have different Hamiltonians and are nontrivial to obtain~\cite{XXZhigher}. We wonder whether our integrable$^*$ phase admits any such extensions, which might shed light on the quasi-integrability of our classical model.

Third, we have only characterized the \textit{boundaries} of the phase diagram in Fig.~\ref{fig:PhaseDiagram}. A straightforward and interesting next step would be to study the quantum-to-classical crossover by better understanding how classical chaos (and perhaps quasi-integrability) at $S\rightarrow\infty$ emerges from the integrable$^*$ regime at $S=1/2$.

In fact, this putative transition between (quantum) integrability and (semiclassical) chaos may also be probed experimentally. The model~\eqref{eq:SeparableModelDelta} can be implemented in a near-term experiment using atomic ensembles confined in a single-mode optical cavity. This would allow for a systematic exploration of the rich physics contained in the dynamical phase diagram (Fig.~\ref{fig:PhaseDiagram}). By changing the local atom density to increase the number of atoms in a given region of constant coupling to the cavity mode, the spin size $S$ can be varied from $S=1/2$ all the way to a semiclassical regime $S\gg 1$: this would enable the experiment to tune between quantum and classical dynamics. Meanwhile, changing the angle between the magnetic field defining the spins' $z$-axis and the axis of the optical cavity allows for tuning of the anisotropy $\Delta$, so that both the special points $\Delta=0,1$ and the regions $\{\Delta < 0\}$, $\{0 < \Delta < 1\}$, and $\{\Delta > 1\}$ can be investigated.

Last, we emphasize that the SVD technique described in Section~\ref{sec:SVDmethod} can be applied directly to the experimental data, revealing the conserved quantities and slow modes. More broadly, we envision using this approach in studying a wider class of physical systems wherein the integrals of motion or their number are not \textit{a priori} known.

In summary, the model~\eqref{eq:SeparableModelDelta} and its associated experimental setup represent a novel paradigmatic platform for studying integrability, chaos, and thermalization under closed many-body quantum dynamics.

\section*{Acknowledgements}
We would like to thank Romain Vasseur, Fabian Essler, Thomas Klein Kvorning, Daniel E. Parker, Emily Davis, and Avikar Periwal for fruitful conversations. This work was supported by the DOE Office of Science, Office of High Energy Physics (GB), the grant DE-SC0019380 (XC, XLQ, MSS, and EA), the ERC synergy grant UQUAM (IDP, XC, and EA), and the Emergent Phenomena in Quantum Systems initiative of the Gordon and Betty Moore Foundation (TS). The numerical computations were carried out on the Lawrencium cluster resource provided by the IT Division at the Lawrence Berkeley National Laboratory under the DOE contract DE-AC02-05CH11231, on the Sherlock computing cluster provided by Stanford University and the Stanford Research Computing Center, and on the cluster of the Laboratoire de Physique Th\'eorique et Mod\`eles Statistiques (CNRS, Universit\'e Paris-Sud).

\bibliography{References.bib}
\begin{widetext}
\renewcommand{\thesection}{S\arabic{section}}    
\renewcommand{\thefigure}{S\arabic{figure}}
\renewcommand{\theequation}{S\arabic{equation}} 

\setcounter{figure}{0}
\setcounter{equation}{0}
\setcounter{section}{0}

\section{Experimental Realization}
\label{appendix:ExperimentalDetails}

\subsection{Derivation of the Effective Hamiltonian}
\label{appendix:ExperimentalH}

Here we elaborate on the derivation of the effective spin Hamiltonian~\eqref{eq:xi_experimental} from the atom-light interaction Hamiltonian~\eqref{eq:ExpIntLowestOrder}, which we now repeat for completeness:
\begin{equation}
H_I \approx \frac{i}{2} \chi \left( \xi_i^* v e^{i\delta t} - \xi_i \adj{v} e^{-i \delta t} \right)\left(\mathbf{S}_i \cdot \hat{c}\right).
\end{equation}
To simplify the following derivation, we will assume that the weights $\xi_i$ are real numbers (although it is interesting to speculate whether one can access an even richer set of dynamics if the weights are allowed to have both non-uniform phases and amplitudes).  In this case, we may write the full Hamiltonian as:
\begin{equation}
    H = \frac{i}{2} \chi \left[ v e^{i\delta t} - \adj{v} e^{-i \delta t} \right] \left[\wS_z \cos \theta + \frac{1}{2} \sin \theta \left( \wS_+ e^{i \omega_Z t} + \wS_- e^{-i \omega_Z t} \right) \right],
    \label{eq:ExpFullH}
\end{equation}
where we have passed into a rotating frame with respect to the atomic Zeeman splitting $\omega_Z$.

Provided the occupation of the $v$ mode remains small, we can adiabatically eliminate it from the dynamics following the approach of Reiter and S\o{}renson~\cite{Reiter2012}. This procedure is essentially a perturbation theory calculation that considers 2-photon scattering processes in which a virtual photon is scattered into the $v$ mode and reabsorbed by the atomic ensemble. Inspecting the Hamiltonian~\eqref{eq:ExpFullH}, we find three distinct processes that add one photon to the $v$ mode:
\begin{align*}
    & -\frac{i}{2} \chi \cos \theta \ \adj{v} \wS_z e^{-i \delta t}, \\
    & -\frac{i}{4} \chi \sin \theta \ \adj{v} \wS_+ e^{- i \delta t + i \omega_Z t}, \\
    & -\frac{i}{4} \chi \sin \theta \ \adj{v} \wS_- e^{- i \delta t - i \omega_Z t}.
\end{align*}
In addition, the Hermitian conjugates of these terms remove a photon from the mode $v$. We must consider all pairs of processes that add a photon to the mode and subsequently reabsorb it. However, only the two-photon processes that are resonant will dominate the slow, effective, ground-state dynamics. For instance, the two-photon process proportional to $v \adj{v} \wS_- \wS_-$ is an off-resonant process and is, therefore, accompanied by a rapidly rotating phase factor $e^{-2 i \omega_Z t}$. As a result, this term quickly averages to zero on timescales $t \gg 1/\omega_Z$, and we are justified in ignoring it. In fact, the only resonant 2-photon processes that survive are the terms proportional to $\wS_z \wS_z, \wS_- \wS_+,$ and $\wS_+ \wS _-$, since all other terms have rapidly oscillating phase factors. The result of this elimination scheme is an effective Hamiltonian for the spins:
\begin{equation}
    H_{\textrm{eff}} = \frac{\chi^2} {4} \left[ \wS_z \wS_z \ \zeta(\delta) \cos^2 \theta + \frac{1}{4} \wS_+ \wS_- \ \zeta(\delta_-) \sin^2 \theta + \frac{1}{4} \wS_- \wS_+ \ \zeta(\delta_+) \sin^2 \theta \right],
    \label{eq:Heff0}
\end{equation}
where $\delta_{\pm} = \delta \mp \omega_Z$ are the detunings from the two-photon resonance, $\kappa$ is the cavity linewidth, and $\zeta(\delta) = \delta/[\delta^2 + (\kappa/2)^2]$.  At large detuning $\delta \gg \kappa, \omega_Z$, Eq.~\ref{eq:Heff0} simplifies to Eq.~\ref{eq:exp_H}, and we obtain the desired model with an anisotropy parameter $\Delta$ controlled by the angle $\theta$ of the magnetic field.

The effective Hamiltonian~\eqref{eq:exp_H}, however, is obtained only if the $\wS_+ \wS_-$ and $\wS_- \wS_+$ terms in Eq.~\ref{eq:Heff0} are balanced, which occurs perfectly only in the limit $\delta \rightarrow \infty$. More generally, at finite $\delta$, the cross-terms $\wS_x \wS_y$ do not cancel and we obtain an effective Hamiltonian
\begin{equation}
    H_{\textrm{eff}} = \frac{\chi^2} {4} \left[ \wS_z \wS_z \ \zeta(\delta) \cos^2 \theta + \frac{1}{4} \left( \wS_x \wS_x + \wS_y \wS_y \right) \ \left(\zeta(\delta_-) + \zeta(\delta_+) \right) \sin^2 \theta \right] + \frac{\chi^2}{4} \sum_i \xi_i^2 S_i^z \left( \zeta(\delta_-) - \zeta(\delta_+) \right),
\end{equation}
which is of the same form as Eq.~\ref{eq:exp_H}, but with additional non-uniform longitudinal magnetic field terms. Such on-site terms, however, are sub-extensive relative to the spin-spin interaction terms. Moreover, we show numerically in Section~\ref{appendix:Commutator} that these additional terms do not affect the integrability of the model at $S = 1/2$ and finite $N$.

\subsection{Effects of Dissipation}
\label{appendix:ExperimentalDissipation}

In addition to the coherent dynamics, the driven cavity system suffers from dissipation due to photon loss from the $v$ mode and atomic free-space scattering from the excited states. These processes can be described formally in a quantum master equation by the relaxation operators
\begin{equation}
    L_v = \sqrt{\kappa} \ v
\end{equation}
and
\begin{equation}
    L_{n,(m,m')} = \sqrt{C_{m,m'}\Gamma_{\mathrm{sc},n}} \left( \ket{m}{\bra{m'}} \right)_n,
\end{equation}
respectively.  Here, $(m,m')$ label internal states of the individual atoms indexed by $n = 1,\ldots,2SN$. $C_{m,m'}$ are branching ratios that sum to unity. $\Gamma_{\mathrm{sc},n} = (\Omega_n / \mathcal{D})^2 \Gamma$ is the free-space scattering rate for the $n$th atom, where $\mathcal{D}$ is the detuning from atomic resonance and $\Gamma$ is the natural excited-state linewidth. In comparison to the scheme presented in Section~\ref{sec:ExpScheme}, note that the Rabi frequency $\Omega_n$ and detuning $\mathcal{D}$ enter likewise into the interaction strength in the Hamiltonian~\eqref{eq:exp_H}, where $(\chi\xi_n)^2 \sim (\Omega_n / \mathcal{D})^2 g_n^2$.  Thus, the scattering rate scales as $\Gamma_{\mathrm{sc},n}\sim (\chi\xi_n)^2/(\eta\kappa)$.

The cavity dissipation described by $L_v$ generates collective dephasing of the atomic ensemble. Upon adiabatically eliminating the cavity mode using the Reiter-S\o{}renson procedure, we find that cavity dissipation leads to an effective relaxation operator
\begin{equation}
    L_{v,\mathrm{eff}} = \sqrt{\kappa} \ \frac{\chi}{2} \left[ \wS_z \frac{\cos \theta}{\delta+i \kappa / 2} + \wS_+ \frac{\sin \theta}{2\delta_+ +i \kappa} e^{i \omega_Z t} + \wS_- \frac{\sin \theta}{2\delta_- +i \kappa} e^{-i \omega_Z t} \right].
\end{equation}
Due to the rapidly oscillating phase factors $e^{\pm i \omega_Z t}$, this relaxation operator may be split into three effective relaxation operators
\begin{subequations}
\begin{align}
    L_{\pm} &= \sqrt{\gamma / 2} \ \mathcal{F}_{\pm}, \\
    L_z &= \sqrt{\Delta \gamma} \ \mathcal{F}_z,
\end{align}
\label{eq:ExpCavityRelax}
\end{subequations}
where we have introduced the collective decay rate
\begin{equation}
\gamma = \frac{\kappa \chi^2 \sin^2\theta}{8[ \delta^2+(\kappa/2)^2]}.
\end{equation}
By contrast, the free-space scattering operators $L_{n,(m,m')}$ act directly on the spins so there is no need to apply the adiabatic elimination procedure.

To what extent do these dissipative processes spoil the signatures of the integrability studied in Section~\ref{sec:ExactDiag}? In the strongly quantum regime at small $S$, for instance, one might be interested in being able to evolve the system long enough to see the plateaus in the autocorrelation function $G_l(t)$, as observed in Fig.~\ref{fig:PlateausDelta05}. To do so, we must ensure that the dissipation timescale is short compared to the timescale $\tau_l$ required for the plateaus to appear. While the full dynamics including both coherent and dissipative processes is difficult to access numerically, we can estimate the dissipative timescales by solving for the dynamics in the presence of dissipation alone. In the absence of coherent dynamics (i.e. $H = 0$), one can compute the equations of motion for the expectation values of operators using the Lindblad equation:
\begin{equation}
    \frac{d}{dt} \left \langle \mathcal{O} \right \rangle = \sum_{r} \left \langle L_{r}^{\dagger} \mathcal{O} L_{r} - \frac{1}{2} L_{r}^{\dagger} L_{r} \mathcal{O} - \frac{1}{2} \mathcal{O} L_{r}^{\dagger} L_{r} \right \rangle,
\end{equation}
where $r$ runs over the set of relaxation operators. Plugging in the free-space scattering operators $L_r = L_{n,(m,m')}$, and assuming an approximately uniform scattering rate $\Gamma_{\mathrm{sc},n} = \Gamma_{\mathrm{sc}}$, one can show that bilinears of spin-$S$ operators $S_j^{\alpha}$ obey the following equations of motion for $j \neq k$:
\begin{subequations}
\begin{align}
    \frac{d}{dt} X_{jk} &= -2(1+C) \Gamma_{\mathrm{sc}} X_{jk}, \\
    \frac{d}{dt} Z_{jk} &= -4 \Gamma_{\mathrm{sc}} Z_{jk},
\end{align}
\end{subequations}
where we have introduced the notation
\begin{align*}
    X_{jk} &= \left \langle S_j^x S_k^x + S_j^y S_k^y \right \rangle, \\
    Z_{jk} &= \left \langle S_j^z S_k^z \right \rangle.
\end{align*}
Thus, the bilinear operators decay on a timescale $\tau \sim 1/\Gamma_{\mathrm{sc}}$ due to free-space scattering.

We can perform a similar calculation for the collective relaxation operators~\eqref{eq:ExpCavityRelax} generated by photon loss from the cavity mode. For large detuning $\delta_{\pm} \approx \delta \gg \kappa,\omega_z$, one can show that the spin bilinears obey the following equations of motion for $j \neq k$:
\begin{subequations}
\begin{align}
    \frac{d}{dt} X_{jk} &= -\gamma \left[\frac{1}{2} \cos^2 \theta \left(\xi_j - \xi_k \right)^2 + \frac{1}{4} \sin^2 \theta \left( \xi_j^2 + \xi_k^2 \right) \right] X_{jk} + 2 \gamma \sin^2 \theta \xi_j \xi_k Z_{jk}, \\
    \frac{d}{dt} Z_{jk} &= - \frac{\gamma}{4} \sin^2 \theta \left[ 2 \left(\xi_j^2 + \xi_k^2 \right) Z_{jk} - \xi_j \xi_k X_{jk} \right].
\end{align}
\end{subequations}
Assuming that the weights $\xi_i$ are numbers of order one, the bilinear operators therefore decay on a timescale $\tau \sim 1/ \gamma$.  A key result of this analysis is that the collective dissipation rate is, somewhat surprisingly, independent of the atom number $\sim NS$, exhibiting no superradiant enhancement. The reason for this is that the average effects of the $L_\pm$ Lindblad operators cancel in the large-detuning limit, where $\delta_+ \approx \delta_-$.  In practice, at finite detuning, any imbalance in these terms can be corrected by adding a weak auxiliary drive field, with no other significant effect on the dynamics.

The relative strengths of decay via the cavity ($\gamma$) and via spontaneous emission ($\Gamma_\mathrm{sc}$) are controlled by the detuning $\delta$.  The total decay rate $\gamma + \Gamma_\mathrm{sc}$, at fixed interaction strength $J^2$, is minimized by choosing a detuning $\delta=\sqrt{1+\eta}\kappa$. For weak to moderate cooperativity $\eta\lesssim 1$, we may instead choose a larger detuning satisfying the conditions $\delta \gg \kappa, \omega_Z$ assumed above.

\section{Quantum model and exact diagonalization results}\label{appendix:EDsubsubsec}

\subsection{Many-body bandwidth}
\label{appendix:Bandwidth}

\begin{center}
\begin{figure}
\includegraphics[width=0.6 \columnwidth]{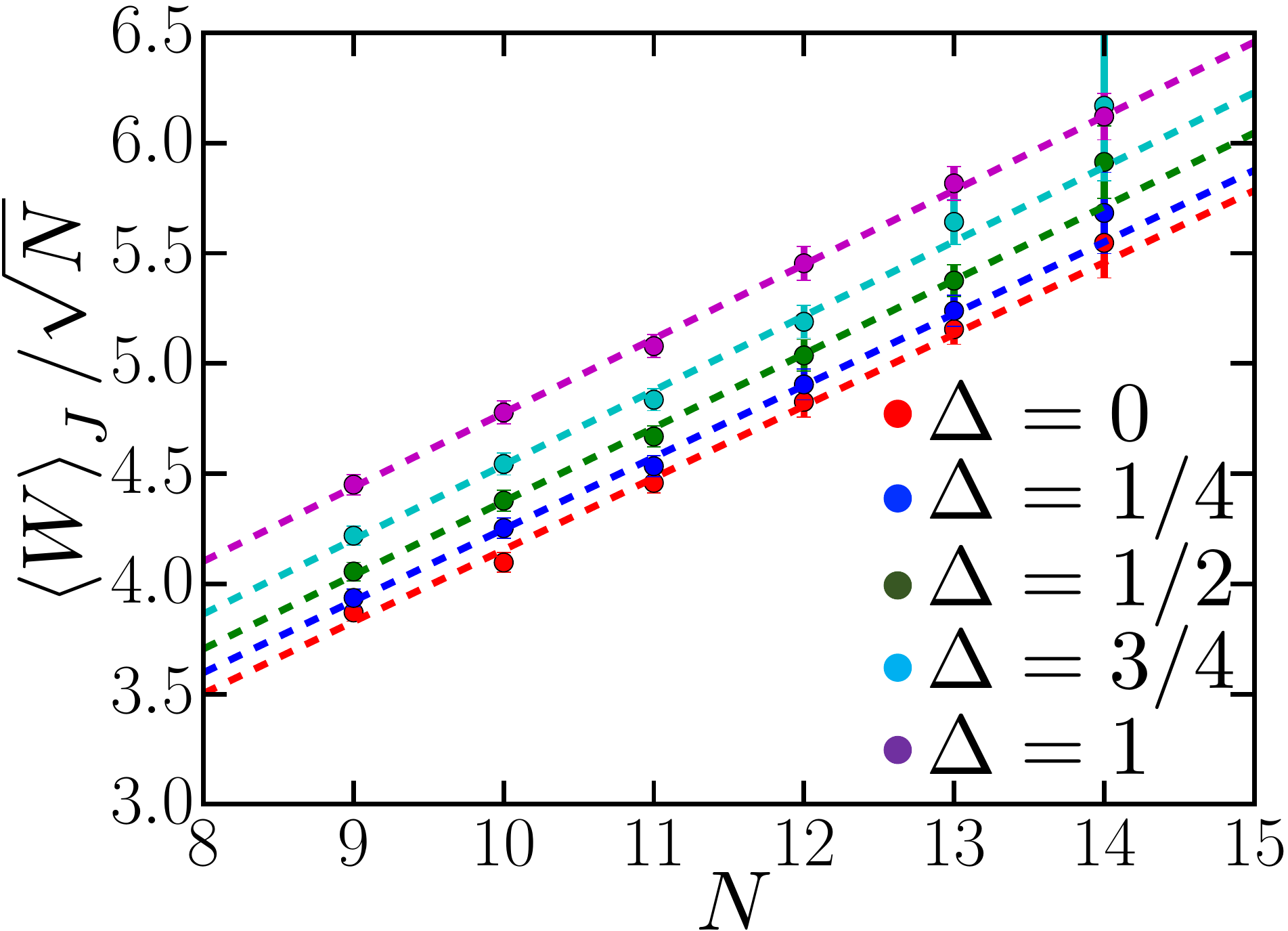}
\caption{Numerically obtained and disorder-averaged many-body bandwidth $\left<{W}\right>_J N^{-1/2}$ as a function of the system size $N$ for different values of the anisotropy $\Delta$ (different colors). The dashed lines correspond to a linear fit. This is consistent with the prediction that $\left<{W}\right>_J \sim \mathcal{O}(N^{3/2})$, as detailed below.}
\label{fig:Bandwidths}
\end{figure}
\end{center}

We now present analytical and numerical evidence that the choice of normalization for the Hamiltonian in Eq.~\ref{eq:SeparableModelDelta} leads to a sensible thermodynamic limit. On the one hand, we show that the many-body bandwidth $W$ scales \emph{superlinearly}, to wit: $W \sim \mathcal{O}(N^{3/2})$. An important caveat is that $W$ \emph{per se} does not mean much unless it is weighed by the many-body density of states $\rho(E)$. On the other hand, the energy fluctuations $\Delta E$ (at least at high temperatures) scale as $\Delta E \sim \mathcal{O}(\sqrt{N})$ which leads to an extensive specific heat, as one would normally expect.

Let us start by rewriting the Hamiltonian as 
\begin{equation}
H = \frac{1}{S\sqrt{N}} \left[\mathcal{F}_x^2 + \mathcal{F}_y^2 + \Delta \mathcal{F}_z^2\right],
\end{equation}
where $\vc{\wS} \equiv \sum_i J_i \mathbf{S}_i$. Each operator $\mathcal{F}_\alpha^2$ has non-negative eigenvalues and we denote the largest one by $f_0^2$. Note that it is the same for all $\alpha \in \{x,y,z\}$ by isotropy. Thus, the largest eigenenergy $E_M$ of $H$ is bounded by
\begin{equation}
E_M \leq \frac{f_0^2}{S\sqrt{N}}(2 + \Delta).
\end{equation}
We can also bound $E_M$ below by using any properly normalized variational state $\ket{\psi}$ since $E_\psi = \bra{\psi} H\ket{\psi} = \sum_n E_n |\langle \psi|n\rangle|^2 \leq E_M \sum_n |\langle \psi|n\rangle|^2  = E_M$. Let us choose
\begin{equation}
    \ket{\psi} = \bigotimes_{i=1}^{N} \ket{S_i^z = \mathrm{sgn}\left(J_i\right) S}
\end{equation}
as a variational state. Then $\bra{\psi} H \ket{\psi} = \frac{\Delta}{S\sqrt{N}}\bra{\psi}\mathcal{F}_z^2\ket{\psi}$ and
\begin{equation}
    \bra{\psi}\mathcal{F}_z^2\ket{\psi} = S^2 \left(\sum_{i=1}^{N} |J_i|\right)^2.
\end{equation}
Moreover, this variational state is an eigenstate of $\mathcal{F}_z$ and has the largest eigenvalue $f_0 = S\sum_i |J_i|$. Thus, we find that
\begin{equation}\label{eq:EMineq}
    \frac{\Delta S}{\sqrt{N}} \left(\sum_{i=1}^{N} |J_i|\right)^2 \leq E_M \leq \frac{(2+\Delta)S}{\sqrt{N}} \left(\sum_{i=1}^{N} |J_i|\right)^2.
\end{equation}
Since the fields $J_i$ are sampled from $\mathcal{N}(0,1)$, the random variable $|J_i|$ has a half-normal distribution with mean $\mu = \sqrt{2/\pi}$ and variance $\sigma^2 = \left(1-\frac{2}{\pi}\right)$. Then $Z \equiv \sum_i |J_i|$ has mean $N\mu$ and variance $N\sigma^2$ so $\left<Z^2\right>_J = N^2\mu^2 + N\sigma^2$. Thus, at large $N$, $ \left< \left(\sum_{i} |J_i|\right)^2\right>_J \approx 2 N^2/\pi$.
After disorder averaging Eq.~\ref{eq:EMineq} we get
\begin{equation}
    \frac{2\Delta S}{\pi} N^{3/2} \lesssim \left< E_M\right>_J \lesssim \frac{2(2+\Delta)S}{\pi} N^{3/2},
\end{equation}
which shows that the disorder averaged $\left< E_M\right>_J \sim \mathcal{O}(N^{3/2})$. 

We now put bounds on the ground state energy $E_m$. Clearly, $E_m \geq 0$. More importantly, we note that
\begin{equation}
    \ket{\phi} = \bigotimes_i \ket{S_i^z = S}
\end{equation}
is an exact eigenstate of $H$ with energy $E_{\phi} = \frac{\Delta S}{\sqrt{N}} \left(\sum_i J_i\right)^2$. Since $\left<\left(\sum_i J_i\right)^2\right>_J = N$, we find that
\begin{equation}
    0 \leq \left<{E_m}\right>_J \leq \left<{E_{\phi}}\right>_J = \Delta S N^{1/2},
\end{equation}
which shows that $\left<{E_m}\right>_J \sim \mathcal{O}(N^{1/2})$. Since $\left< E_M \right>_J \sim \mathcal{O}(N^{3/2})$, this concludes our proof that the disorder-averaged bandwidth scales as $\left<{W}\right>_J \sim N^{3/2}$. We have also computed $\left<{W}\right>_J$ numerically and we find an excellent agreement with the above analysis, as shown in Fig.~\ref{fig:Bandwidths}.

Finally, we note that although $W$ scales superlinearly, it is quite possible that the bulk of the many-body states are located around a typical energy $E_{\mathrm{typ}} \sim \mathcal{O}(N)$ and that the density of states has a long and thin tail extending all the way to $\mathcal{O}(N^{3/2})$.

\subsection{Energy fluctuations}
\label{appendix:Fluctuations}

\begin{center}
\begin{figure}
\includegraphics[width=0.6 \columnwidth]{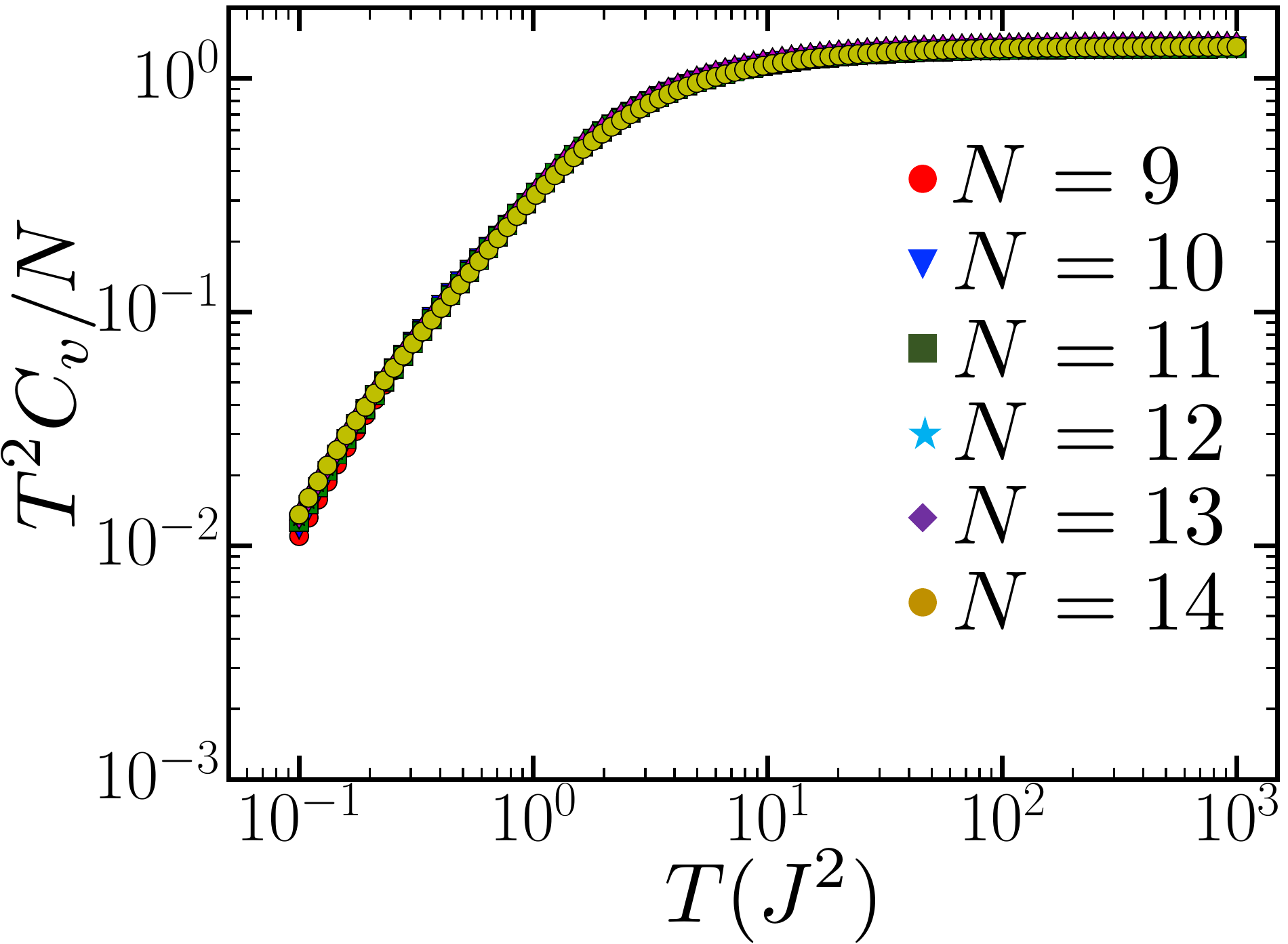}
\caption{Numerically obtained scaling collapse of $C_vT^2/N$ (where $C_v$ is the specific heat) as a function of the temperature $T$. We see that at high temperatures the specific heat approaches $\frac{c}{T^2} N$ with $c\approx 1.4$. This is in very good agreement with the result from Eq.~\ref{eq:InfiniteTSpecificHeat} wherein the constant is $3/2$.}
\label{fig:SpecificHeat}
\end{figure}
\end{center}

We now show that the disorder-averaged thermal energy fluctuations ${(\Delta E)^2}$ at high temperatures scale as $\left<(\Delta E)^2\right>_J  = c N$ for some constant $c$. Note that, by the fluctuation-dissipation theorem, this means that the specific heat \emph{per particle} behaves as $\frac{C_v}{N} \approx \frac{c}{T^2}$ at high temperatures $T$.

The energy fluctuations are defined as
\begin{equation}
    \left(\Delta E\right)^2 = \langle H^2\rangle_{T} - \langle H\rangle_{T}^{2},
\end{equation}
where $\langle\dots\rangle_T \equiv \left(\Tr e^{-\beta H}\right)^{-1} \Tr\left[e^{-\beta H}\dots\right]$ denotes thermal averaging. For simplicity, we will focus on the $S=1/2$ and $\Delta=1$  case. At lowest order in $\beta$ we can approximate
\begin{eqnarray}
    \langle H\rangle_T &\approx& \frac{1}{2^N} \frac{3}{S\sqrt{N}} \sum_{ij} J_i J_j \Tr\left[S_{i}^{\alpha}S_{j}^{\alpha}\right] \\ \nonumber
    &=& \frac{3}{2\sqrt{N}} \left(\sum_{i} J_i^2\right) \\ \nonumber
    &=& \frac{3}{2\sqrt{N}} Z,
\end{eqnarray}
where $Z \sim \chi^2(N)$. Similarly, the first term becomes
\begin{eqnarray}
    \langle H^2\rangle_T &\approx& \frac{1}{2^N}\frac{1}{S^2N} \sum_{ijkl} J_{i}J_{j}J_{k}J_{l} \sum_{\alpha\beta} \Tr\left[S_{i}^{\alpha}S_{j}^{\alpha}S_{k}^{\beta}S_{l}^{\beta}\right] \\ \nonumber
    &=& \frac{1}{N 2^{N-2}}  \sum_{ijkl} J_{i}J_{j}J_{k}J_{l} \left\{\sum_{\alpha}  \Tr\left[S_{i}^{\alpha}S_{j}^{\alpha}S_{k}^{\alpha}S_{l}^{\alpha}\right]  + \sum_{\alpha}\sum_{\beta \neq \alpha} \Tr\left[S_{i}^{\alpha}S_{j}^{\alpha}\right]\Tr\left[S_{k}^{\beta}S_{l}^{\beta}\right]\right\} \\ \nonumber
    &=& \frac{1}{N}  \sum_{ijkl} J_{i}J_{j}J_{k}J_{l} \left\{9\left[\Tr\left(S_{i}^{z}\right)^2\right]^2 \delta_{ij}\delta_{kl} + 6\left[\Tr\left(S_{i}^{z}\right)^2\right]^2\delta_{ij}\delta_{kl}\right\} \\ \nonumber
    &=& \frac{15}{4N} \left(\sum_{i} J_i^2\right)^2 \\ \nonumber
    &=& \frac{15}{4N} Z^2.
\end{eqnarray}
Putting the two results together we find that $\left(\Delta E\right)^2 \approx \frac{3}{2N} Z^2$. Since $Z$ is chi-squared distributed $\chi^2(N)$, we know that $\mathbb{E}\left[Z^2\right] = N^2 + 2N$. Thus, after averaging over disorder, we get
\begin{equation}\label{eq:InfiniteTSpecificHeat}
    \left<\left(\Delta E\right)^2\right>_J \approx \frac{3}{2} N.
\end{equation}
This ensures that the specific heat $C_{v} = \frac{3}{2T^2} N$ is extensive. We can also compute this quantity numerically using the exact spectrum of $H$ and we find an excellent agreement, as shown in Fig.~\ref{fig:SpecificHeat}.

\begin{center}
\begin{figure}
\includegraphics[width= \columnwidth]{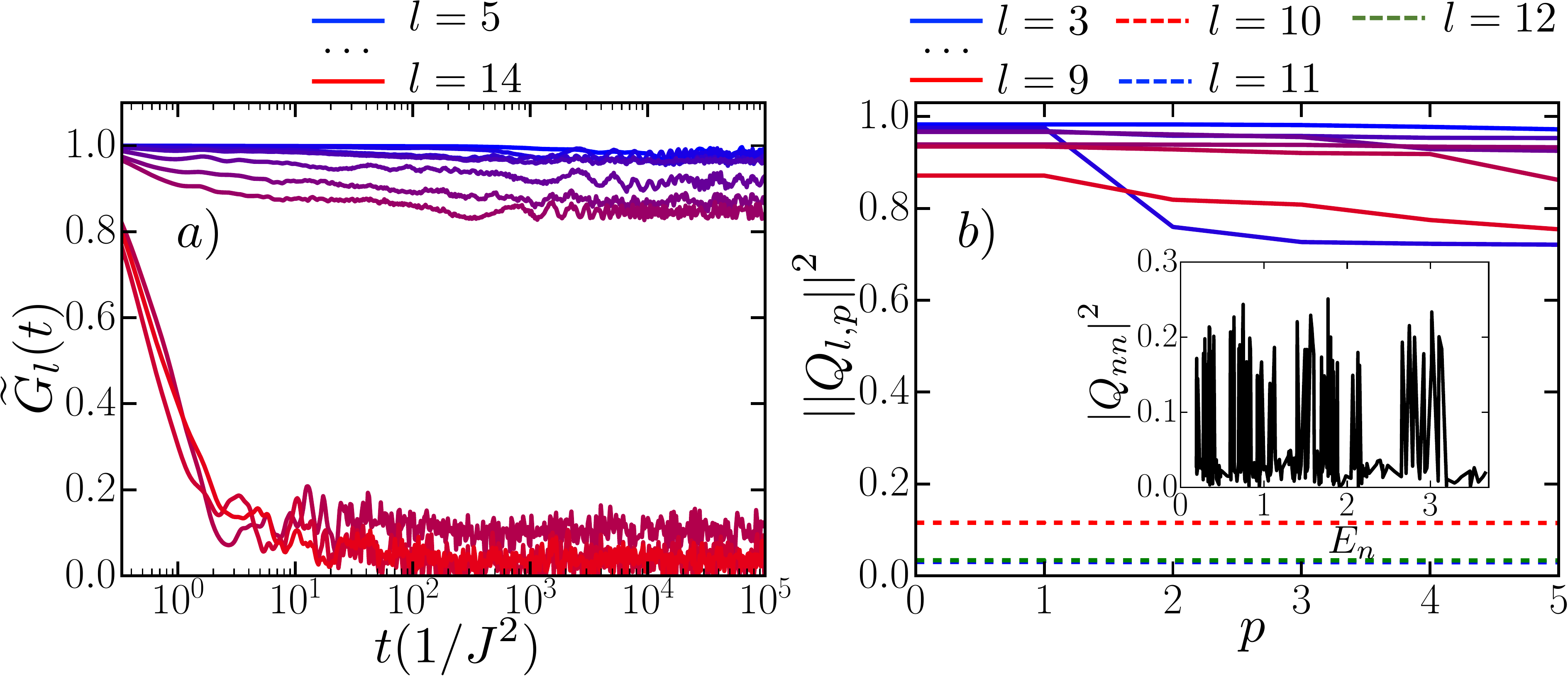}
\caption{Effects due to the overlap between the slow modes $\hat{Q}_l$ and higher powers of the exactly conserved quantities, $\hat{H}$ and $\hat{S}_\mathrm{tot}^z$. These results correspond to a fixed disorder realization for the $\{J_i\}$ in a system of $N=9$ spins ($S=1/2$) with $\Delta = 0.9$.\\
a) The correlation function $\widetilde{G}_l(t)$ for the slow modes $\widetilde{Q}_l$ with $l>4$ in a space orthogonal to the one spanned by the operators defined in Eq.~\ref{eq:OperatorsHigherPowers}. We use a color gradient between $l=5$ (blue curve) and $l=14$ (red curve).\\
b) The plateau values after we substract the contributions from $\hat{H}^{p}$ as a function of the power $p$ (see the definition in Eq.~\ref{eq:QlpOverlapMeasure}). We use a color gradient for the solid curves in which $l=3$ corresponds to the blue curve and $l=9$ corresponds to the red curve; the red, blue, green dashed curves correspond to $l=10,11,12$, respectively. (inset) The diagonal matrix elements $\left|\bra{n}\hat{Q}_3\ket{n}\right|^2$ for the mode $l=3$ as a function of the energy eigenvalue $E_n$. This example provides further evidence that the slow modes $\hat{Q}_l$ are far away from being projectors on many-body energy eigenstates.}
\label{fig:Overlaps}
\end{figure}
\end{center}

\subsection{Overlaps with higher powers of the conserved quantities $\hat{H}$ and $\hat{S}_{\mathrm{tot}}^z$}
\label{appendix:OverlapH2}

We now address the possibility that the large plateau values (Fig.~\ref{fig:Plateaus}) observed in the quantum model at intermediate $\Delta \in (0,1)$ are caused by a significant overlap between the slow operators $\hat{Q}_l$ and higher powers of the conserved quantities, $\hat{H}$ and $\hat{S}_{\mathrm{tot}}^z$. We use a twofold approach, as detailed below.

Let us first summarize why this is a legitimate concern. Recall that at $\Delta \in (0,1)$, the SVD analysis finds two exactly conserved spin bilinears, $\hat{H}$ and $(\hat{S}_{\mathrm{tot}}^z)^{2}$, along with another $N-2$ slow modes $Q_l$, where $l = 3, \dots, N$. The auto-correlation function of the latter operators exhibits a large long-time plateau, which we interpreted as a consequence of the existence of further conservation laws~\eqref{eq:qiom} and of quantum integrability. However, it is possible the the $\hat{Q}_l$s overlap non-trivially with exactly conserved operators such as $\hat{H}^{p} (\hat{S}_{\mathrm{tot}}^z)^{p'}$: these higher powers of the exactly conserved quantities consist of $k$-body terms, including $k=2$, so they have non-zero support in the space of spin bilinears and, thus, can conceivably overlap with $\hat{Q}_l$.

Furthermore, the support in the space of $2$-local operators can be non-zero even in the thermodynamic limit $N\rightarrow \infty$ due to the all-to-all nature of the interaction. For instance, consider $\hat{\mathcal{O}}_1 = \frac{1}{\sqrt{N}}\hat{H}$. By the previous section, we know that $\frac{1}{\Tr\left[\mathbf{1}\right]}\Tr\left[\hat{\mathcal{O}}_1^2\right]$ is of order one even in the thermodynamic limit. Similarly, let $\hat{\mathcal{O}}_2 = \frac{1}{N} \left(\hat{S}_{\mathrm{tot}}^z\right)^2$ such that $\frac{1}{\Tr\left[\mathbf{1}\right]}\Tr\left[\hat{\mathcal{O}}_2^2\right]$ is also of order one as $N \rightarrow \infty$. It is not hard to check that the only $2$-body terms emerging from higher powers of the conserved quantities, namely operators of the form $\hat{\mathcal{O}}_1^{p}\hat{\mathcal{O}}_2^{p'}$, that do \emph{not} vanish in the thermodynamic limit are:
\begin{eqnarray}\label{eq:OperatorsHigherPowers}
    \hat{\mathcal{O}}_1 &=& \frac{1}{\sqrt{N}}\hat{H}, \\ \nonumber
    \hat{\mathcal{O}}_2 &=& \frac{1}{N}\left(\hat{S}_{\mathrm{tot}}^z\right)^2, \\ \nonumber
    \hat{\mathcal{O}}_3 &=& \frac{1}{N}\sum_{ij} J_i J_j \hat{S}_i^z \hat{S}_j^z, \\ \nonumber
    \hat{\mathcal{O}}_4 &=& \frac{1}{N}\sum_{ij} J_i \hat{S}_i^z \hat{S}_j^z. \nonumber
\end{eqnarray}
Above, $ \hat{\mathcal{O}}_3$ can be generated by $\hat{\mathcal{O}}_1 \hat{\mathcal{O}}_1$ as long as $\Delta \in (0,1)$, and $ \hat{\mathcal{O}}_4$ by $\hat{\mathcal{O}}_1 \hat{\mathcal{O}}_2$ (in fact, one gets a prefactor $\frac{1}{N}\sum_k J_k$; it scales as $1/\sqrt{N}$ typically if $J_k$ has zero mean and as $O(1)$ if $\overline{J_k} \neq 0$; by precaution, we still include it in the following analysis).  
We perform a Gram-Schmidt orthonormalization of these four operators to obtain a basis set $\left\{\widetilde{\mathcal{O}}_1,\widetilde{\mathcal{O}}_2,\widetilde{\mathcal{O}}_3,\widetilde{\mathcal{O}}_4\right\}$ wherein $\frac{1}{\Tr\left[\mathbf{1}\right]}\Tr\left[\widetilde{\mathcal{O}}_i\widetilde{\mathcal{O}}_j\right] = \delta_{ij}$. We then define a matrix $R$ whose columns vectors are the $\widetilde{\mathcal{O}}_i$s. This allows us to perform the SVD analysis in the operator space orthogonal to the one spanned by the $\widetilde{\mathcal{O}}_i$s: using the time series matrix $M$ from Eq.~\ref{eq:defM}, we perform the SVD decomposition of $\left(\mathbf{1} - RR^{\dagger}\right)M$ instead. Trivially, we find that the first four singular values are exactly zero so we focus on the modes corresponding to $l>4$. Following the prescription from Section~\ref{subsec:ModesDecayQuantum}, we define the slow operators $\widetilde{Q}_l$ for $l\geq 5$ and compute their auto-correlation function $\widetilde{G}_l(t) = \frac{1}{\Tr\left[\mathbf{1}\right]} \Tr\left[\widetilde{Q}_l(t)\widetilde{Q}_l(0)\right]$, as shown in Fig.~\ref{fig:Overlaps}a. Thus, we see that the overlap with higher powers of $\hat{H}$ and $\hat{S}_{\mathrm{tot}}^z$ has a negligible effect on the plateau values of the slow modes $\hat{Q}_l$.

In the second approach, we numerically study the effect of $\hat{H}^{p}$ on the plateau values by analyzing the matrix elements $Q_{l,nn} \equiv \bra{n} \hat{Q}_l\ket{n}$, where $\ket{n}$ is an energy eigenstate, $H\ket{n} = E_n \ket{n}$. Let us define the vector $\vec{q}_l$ whose elements are $q_{l}^{(n)} = \frac{1}{\sqrt{\Tr\left[\mathbf{1}\right]}} \left|Q_{l,nn}\right|$. It is known that the norm of this vector is just the plateau value of the $Q_l$ mode, whenever the latter is well-defined: $g_l = \norm{\vec{q}_l}^2$. Another way of studying the effect of $\hat{H}^{p}$ on the plateau value is to analyze the overlap between $\vec{q}_l$ and the set of vectors $\left\{\vec{E}^{(p)}: E_{n}^{(p)} = \left(E_{n}\right)^{p}, 0\leq p \leq P\right\}$, i.e. the powers of the energy eigenvalues.

Once again, we perform a Gram-Schmidt orthonormalization of the above vectors to obtain a set $\left\{\vec{\eta}_i: 0\leq i \leq P\right\}$ wherein $\langle \vec{\eta}_i |\vec{\eta}_j\rangle \equiv \vec{\eta}_i^{T} \cdot \vec{\eta}_j = \delta_{ij}$. We then define
\begin{equation}\label{eq:QlpOverlapMeasure}
    \norm{Q_{l,p}}^2 = \norm{\vec{q}_l}^2 - \sum_{i=1}^{p} \left|\vec{q}_l^{T} \cdot \vec{\eta}_i\right|^2,
\end{equation}
which is a measure of the residual plateau value for the mode $\hat{Q}_l$ after removing the contributions of $\hat{H}^{0},\hat{H}^{1}, \dots, \hat{H}^p$. Naturally, for any fixed mode index $l$, $\norm{Q_{l,p}}^2$ decreases from the plateau value $\norm{Q_{l,0}}^2 = \norm{\vec{q}_l}^2 = g_l$ as $p$ increases. Numerically, in Fig.~\ref{fig:Overlaps}b (main plot), we observe that for the slow modes, $2< l < N$, the decrease is small compared to $g_l$ for powers up to $p \sim N/2$, at which point $\hat{H}^p$ becomes an $N$-body operator. In fact, $\norm{Q_{l,p}}^2$ depends roughly linearly on $p$ and decays approximately to zero only when $p \sim 2^N\gg N$. We have also checked that the result is not altered upon including low powers of $ \hat{S}^z_{\mathrm{tot}}$ (i.e. removing contributions from $ \hat{H}^{p} (\hat{S}^z_{\mathrm{tot}})^{r}$, for $p, r \le N$). Therefore, the plateaus of the slow modes are not caused by simple functions of the exactly conserved quantities, $\hat{H}$ and $\hat{S}^z_{\mathrm{tot}}$, at $\Delta \in (0,1)$. Finally, the inset of Fig.~\ref{fig:Overlaps}b shows that slow modes $\hat{Q}_l$ are not a linear combination of a few projectors onto energy eigenstates. 

To summarize, these analyses render highly unlikely the possibility that the slow mode plateaus at intermediate $\Delta$ are solely due to the overlap with the known conserved quantities. This reinforces the thesis of the existence of further \textit{nontrivial} conserved quantities~\eqref{eq:qiom}, as put forth in the main text.

\begin{center}
\begin{figure}
\includegraphics[width= 0.6\columnwidth]{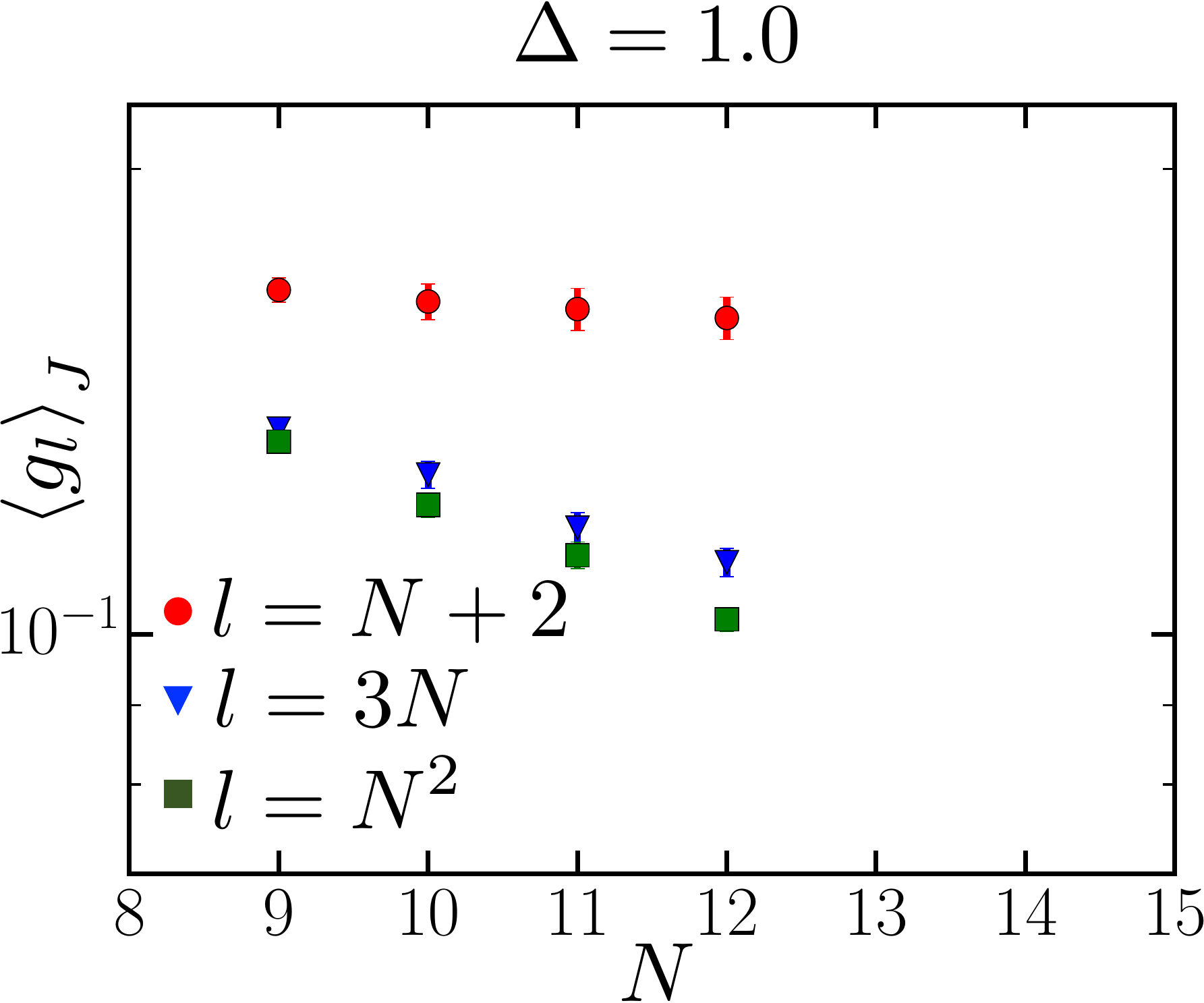}
\caption{Semilog plot of of the disorder-averaged plateau values $\left< g_{l}\right>_{J} = \left< G_l\left(t\rightarrow \infty\right)\right>_{J}$ as a function of $N$ for three modes in the middle of the SVD spectrum at $\Delta=1.0$: the $N+2$ mode which is expected to be the first decaying mode, the $3N$ mode, and the $3N^2$ mode. We see that the latter two decay as $N$ increases, but this is not completely clear for the $N+2$ mode.}
\label{fig:MiddleSVDspectrum}
\end{figure}
\end{center}

\subsection{The middle of the SVD spectrum}
\label{appendix:MiddleSVD}

We now check that, at $\Delta=1.0$, we do not find more than $N+1$ bilinear conserved quantities, as suggested by the analytic results of Section~\ref{sec:IntegrablePoints}. More precisely, we numerically compute the plateau values corresponding to the $l=N+2$, $l=3N$, and $l=N^2$ modes obtained from the quantum SVD analysis. As shown in Fig.~\ref{fig:MiddleSVDspectrum}, we see that the latter two plateau values decrease with the system size $N$. However, this is not entirely clear for the $N+2$ mode as its flow with system size is inconclusive---we conjecture that this is a finite size effect rather than a sign of an additional conserved quantity.

\subsection{Level statistics}
\label{appendix:LevelStatistics}

We also analyze the level statistics of the quantum Hamiltonian~\eqref{eq:SeparableModelDelta} by studying the ratio of adjacent energy levels $E_n$ defined by $r_n = \min\{\Delta E_{n},\Delta E_{n+1}\}/\max\{\Delta E_{n},\Delta E_{n+1}\}$, where $\Delta E_{n} = E_{n+1} -E_{n}$. We now provide details vis-\`a-vis the various symmetries and subsequent degeneracies we have to take into account.

At $\Delta=1$ the Hamiltonian has the full $\mathrm{SU}(2)$ symmetry and the eigenstates of $H$ are grouped in blocks corresponding to irreducible representations (irreps). Each state is simultaneously an eigenstate of $\hat{S}_{\mathrm{tot}}^2$ with an eigenvalue $S_{\mathrm{tot}}(S_{\mathrm{tot}}+1)$ and of $\hat{S}_{\mathrm{tot}}^z$ with an eingenvalue $S_{\mathrm{tot}}^{z}\in \{-S_{\mathrm{tot}},\dots,S_{\mathrm{tot}}\}$. Thus, each block contains $2S_{\mathrm{tot}}+1$ eigenstates of $H$. We keep the states in a fixed sector of both $\hat{S}_{\mathrm{tot}}^2$ and $\hat{S}_{\mathrm{tot}}^z$ by selecting the block $S_{\mathrm{tot}}$ that has the largest multiplicity in the irreps. decomposition (i.e. the most frequent block to obtain the largest number of viable states). 

At $\Delta \neq 1$ the symmetry is $\mathrm{U}(1)\rtimes\mathcal{T}$: $\mathrm{U}(1)$ is generated by $U_{\phi} = \exp\left(-i\phi S_{\mathrm{tot}}^z\right)$ and it corresponds to the conservation of $S_{\mathrm{tot}}^z$, whereas $\mathcal{T}$ represents the anti-unitary time reversal symmetry. Moreover, there is a mirror symmetry corresponding to $\hat{S}_z \rightarrow -\hat{S}_{z}$. For the cases wherein $\mathcal{T}^{2} = -1$, we have Kramers doublets and we keep the states solely in the $S_\mathrm{tot}^z = 1/2$ sector because it is the most populous. For the cases where $\mathcal{T}^2 = +1$, we keep states solely in the $S_{\mathrm{tot}}^z = 1$ sector in order to account for the mirror symmetry as well (the states in the $S_{\mathrm{tot}}^z = 0$ sector are degenerate under the mirror symmetry).

\begin{center}
\begin{figure}
\includegraphics[width=\columnwidth]{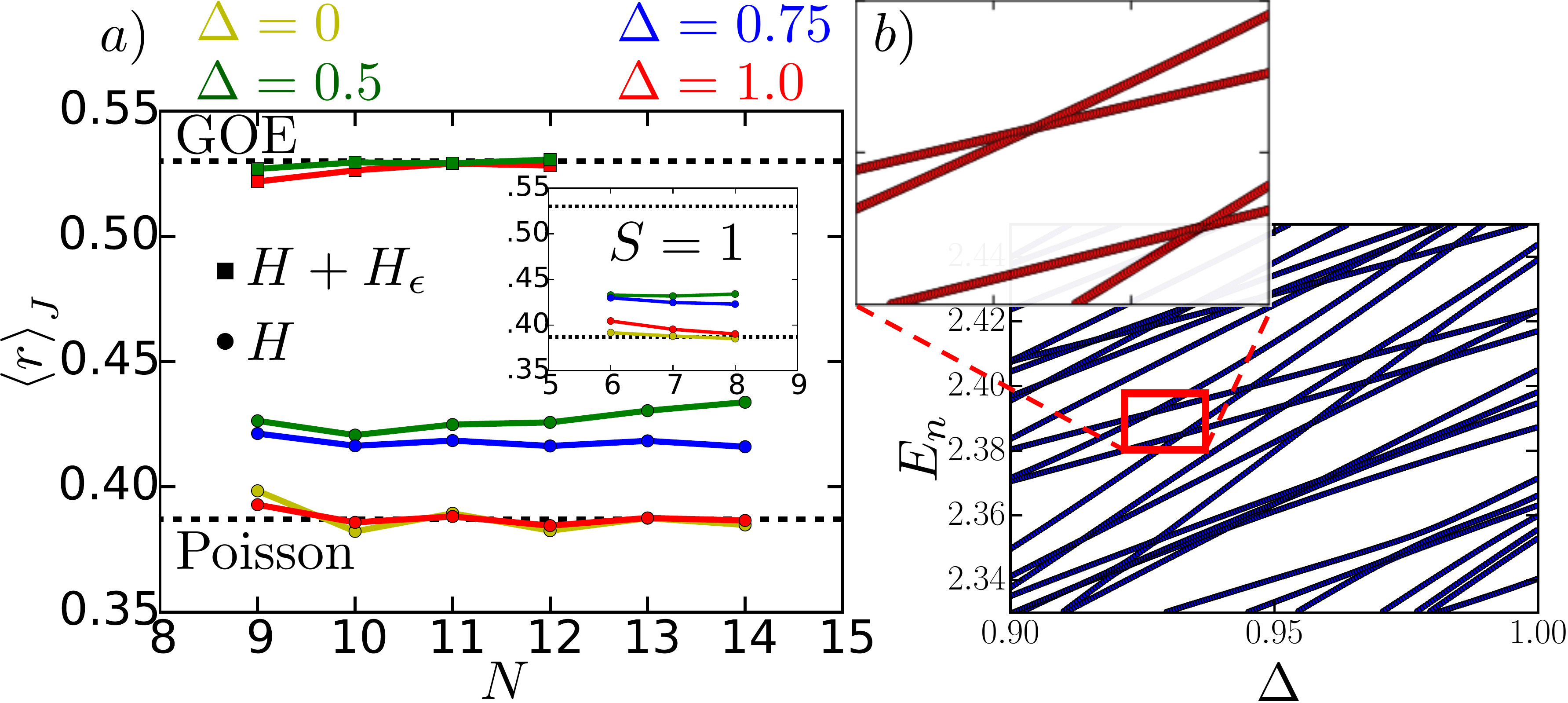}
\caption{a) Plot of the disorder-averaged $\left< r\right>_{J}$ as a function of the system size $N$ for $S=1/2$. Different colors correspond to different values of the anisotropy: red for $\Delta=1$, blue for $\Delta=0.75$, green for $\Delta=0.5$, and yellow for $\Delta=0$. We see that, for the Hamiltonian $H$ from Eq.~\ref{eq:SeparableModelDelta}, the level statistics is close to Poisson (the round markers) as it would be for an integrable model. Adding the perturbation $H^{(1)}_{\epsilon=0.1}$ from Eq.~\ref{eq:ChaoticPerturbation} renders the level statistics of $H+H^{(1)}_{\epsilon}$ close to GOE (the square markers), signaling level repulsion and a chaotic behavior. (inset) We obtain qualitatively similar results for a Hamiltonian~\eqref{eq:SeparableModelDelta} consisting of $S=1$ degrees of freedom. \\
b) Energy levels of $H$ within a fixed $S_{\mathrm{tot}}^{z}=1/2$ sector as a function of $\Delta$ in a single disorder realization $\{J_i\}$ for a system of $N=11$ spins ($S=1/2$). We see many ``real'' (i.e. not avoided) level crossings, as shown in the inset which is a zoom on two such events. This violation of the Wigner-von Neumann non-crossing rule provides further evidence of quantum integrability at intermediate $0<\Delta<1$.}
\label{fig:Rratios}
\end{figure}
\end{center}

Having selected the eigenstates of $H$ to account for the degeneracies due to symmetries, we compute the ratio $r_n$ for the adjacent levels. We then average over all $r_n$ and over disorder realizations to obtain $\left< r\right>_{J}$. We know that a chaotic system exhibits level repulsion and has GOE level statistics characterized by a value $\langle r \rangle_{\mathrm{GOE}} \approx 0.53$. An integrable system does not have level repulsion (its energy levels are uncorrelated) and generically has Poissonian level statistics characterized by a value $\langle r \rangle_{\mathrm{Poisson}} \approx 0.387$.

In Fig.~\ref{fig:Rratios}a we plot $\left< r\right>_{J}(N)$ for different values of the anisotropy $\Delta$. We find that both $\Delta = 0$ and $\Delta =1$ exhibit almost perfectly Poissonian level statistics. However, for intermediate $0<\Delta<1$, the system of $S=1/2$ spins (or $S=1$ for the inset) is not as close to Poisson, although it is even farther away from GOE. Moreover, we cannot extract a meaningful flow with the system size $N$ towards either GOE or Poisson. We note, in passing, that there are exceptions from the equivalence between integrability and Poissonian level statistics: as detailed in Ref.~\onlinecite{Caux} (see also the references therein), there are well-known counterexamples~\cite{Barba2008,BerryTabor,Schliemann2010,Yuzbashyan2016} of integrable systems whose level statistics are neither Poisson nor GOE.

Nonetheless, if we add the perturbation $H_{\epsilon}^{(1)}$ from Eq.~\ref{eq:ChaoticPerturbation} to $H$ we immediately see that the system obeys almost perfectly GOE statistics even for rather small system sizes (see the curves with square markers in Fig.~\ref{fig:Rratios}a).

In Fig.~\ref{fig:Rratios}b we work in a fixed disorder realization for the fields $\{J_i\}$ and we adiabatically change the anisotropy $\Delta$. Then we plot the exact energy levels within a fixed symmetry sector, as detailed above, and explicitly check whether there exist level crossings or if they are avoided (level repulsion). We find that there are numerous real crossings at intermediate $\Delta$, which provides further qualitative evidence that $H$ is integrable for $S=1/2$.

\begin{center}
\begin{figure}
\includegraphics[width= 0.6\columnwidth]{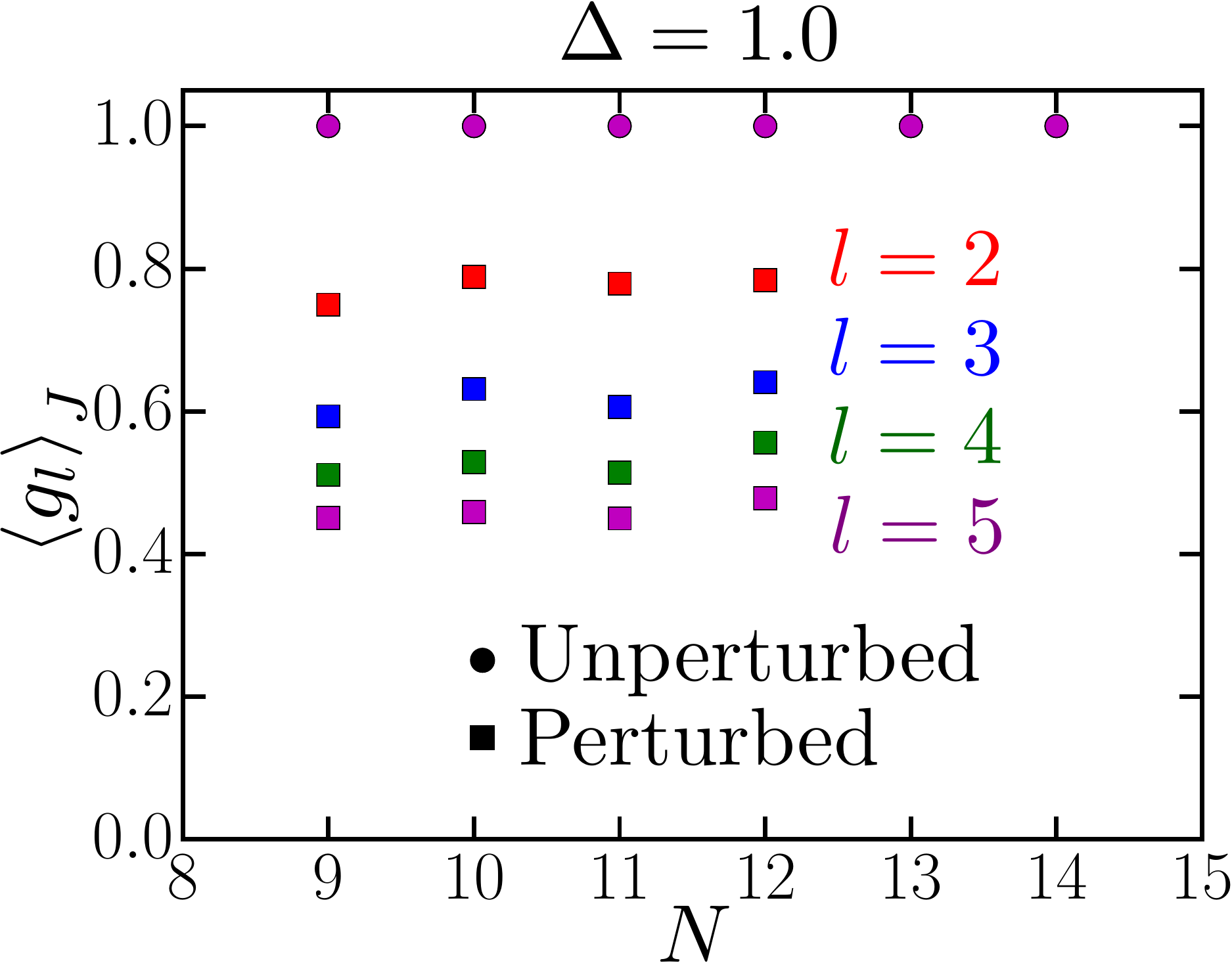}
\caption{Plot of the disorder-averaged plateau values $\left< g_{l}\right>_{J} = \left< G_l\left(t\rightarrow \infty\right)\right>_{J}$ as a function of the system size $N$ at $\Delta=1.0$. The overlapping circular markers correspond to the lowest four modes of the unperturbed model~\eqref{eq:SeparableModelDelta}, which has $N+1$ bilinear conserved quantities. The square markers correspond to adding a perturbation $\frac{1}{S\sqrt{N}}\sum_i J_i^2 S_{i}^z$ and the different colors correspond to the lowest four modes (that are not exactly conserved) in increasing order of their singular values: red, blue, green, and magenta. While the perturbation decreases the plateau value, we note that there is no flow with system size: the perturbation is sub-extensive (suppressed by $1/\sqrt{N}$).}
\label{fig:CorrelatedFields}
\end{figure}
\end{center}

\subsection{Stability to the commutator perturbation}
\label{appendix:Commutator}

We conclude by studying the effect of the perturbation~\eqref{eq:FlipFlopperturbation} that arises upon recasting the Hamiltonian into the ``flip-flop'' form of Eq.~\ref{eq:FlipFlopH}. The term $J_i J_j S_i^{+}S_{j}^{-}$ from~\eqref{eq:FlipFlopH} in the main text can be written as
\begin{eqnarray}
    S_{i}^{+}S_{j}^{-} &=& \left(S_{i}^x + i S_{i}^y\right)\left(S_{j}^x - i S_{j}^y\right) \\ \nonumber
    &=& S_{i}^xS_{j}^x + S_i^yS_j^y + i\left[S_i^y,S_j^x\right] \\ \nonumber
    &=& S_{i}^xS_{j}^x + S_i^yS_j^y + S_i^z \delta_{ij},
\end{eqnarray}
which leads an overall term $\frac{1}{S\sqrt{N}}\sum_i J_i^2 S_i^z$ in addition to the Hamiltonian $H$ defined in Eq.~\ref{eq:SeparableModelDelta}. We have argued that this is an irrelevant perturbation in a thermodynamic system with $N \gg 1$ degrees of freedom. 

We now explicitly study its impact on the plateau values of the auto-correlation function $G_l(t)$. As shown in Fig.~\ref{fig:CorrelatedFields}, we find that the effect of this perturbation is only quantitative: the modes reach plateau values that are lower than those occuring in the unperturbed model, but they do \emph{not} decay as the system size $N$ increases. We note that the quantitative shift is also due to the fact that the slow modes we compute are bilinear operators even though the perturbation $\frac{1}{S\sqrt{N}}\sum_i J_i^2 S_i^z$ is a linear combination of $1$-body terms.

\end{widetext}

    \end{document}